\documentclass{aa}  

\usepackage{graphicx}
\usepackage{xcolor}
\usepackage{float}
\usepackage{txfonts}
\usepackage{subfig}
\usepackage{hyperref} 
\hypersetup{colorlinks=true,citecolor=blue}
\usepackage[normalem]{ulem}

\begin{document}

   \title{Spin evolution of Venus-like planets subjected \\ to gravitational and thermal tides}


   \author{A. Revol\inst{1}
          \and
          E. Bolmont\inst{1}
          \and
          G. Tobie\inst{2}
          \and
          C. Dumoulin\inst{2}
          \and
          Y. Musseau\inst{2}
          \and
          S. Mathis\inst{3}
          \and
          A. Strugarek\inst{3}
          \and
          A.S. Brun\inst{3}
          }
   \institute{Observatoire de Genève, Centre pour la Vie dans l'Univers, Université de Genève,
              51 Chemin Pegasis, 1290 Sauverny, Switzerland.\\
              \email{alexandre.revol@unige.ch, emeline.bolmont@unige.ch}
         \and
             Laboratoire de Planétologie et Géosciences, UMR-CNRS 6112, Nantes Université, 2 rue de la Houssinière, BP 92208, 44322 Nantes Cedex 3, France
        \and
            AIM, CEA, CNRS, Université Paris-Saclay, Université Paris Diderot, Sorbonne Paris Cité, 91191 Gif-sur-Yvette Cedex, France.
             }

   \date{Received 26 December 2022 / Accepted 28 February 2023}


  \abstract
   {The arrival of powerful instruments will provide valuable data for the characterization of rocky exoplanets.
   Rocky planets are mostly found in close-in orbits.
   They are therefore usually close to the circular-coplanar orbital state and are thus considered to be in a tidally locked synchronous spin state.
   For planets with larger orbits, however, exoplanets should still have nonzero eccentricities and/or obliquities, and realistic models of tides for rocky planets can allow for higher spin states than the synchronization state in the presence of eccentricities or obliquities.
   }
   {
   This work explores the secular evolution of a star-planet system under tidal interactions, both gravitational and thermal, induced by the quadrupolar component of the gravitational potential and the irradiation of the planetary surface, respectively.
   We show the possible spin-orbit evolution and resonances for eccentric orbits and explore the possibility of spin-orbit resonances raised by the obliquity of the planet.
   Then, we focus on the additional effect of a thick atmosphere on the possible resulting spin equilibrium states and explore the effect of the evolution of the stellar luminosity.
   }
   {
   We implemented the general secular evolution equations of tidal interactions in the secular code called ESPEM.
   In particular, we focus here on the tides raised by a star on a rocky planet and consider the effect of the presence of an atmosphere, neglecting the contribution of the stellar tide.
   The solid part of the tides was modeled with an anelastic rheology (Andrade model), while the atmospheric tides were modeled with an analytical formulation that was fit using a global climate model simulation.
   We focused on a Sun-Venus-like system in terms of stellar parameters, orbital configuration and planet size and mass.
   The Sun-Venus system is a good laboratory for studying and comparing the possible effect of atmospheric tides, and thus to explore the possible spin state of potential Venus-like exoplanets.
   }
   {
   The formalism of Kaula associated with an Andrade rheology allows spin orbit resonances on pure rocky worlds.
   Similarly to the high-order spin-orbit resonances induced by eccentricity, the spin obliquity allows the excitation of high-frequency Fourier modes that allow some spin-orbit resonances to be stable.
   If the planet has a dense atmosphere, like that of Venus, another mechanism, the thermal tides, can counterbalance the effect of the gravitational tides.
   We found that thermal tides change the evolution of the spin of the planet, including the capture in spin-orbit resonances.
   If the spin inclination is high enough, thermal tides can drive the spin toward an anti-synchronization state, that is, a the 1:1 spin-orbit resonance with an obliquity of 180~degrees.
   }
   {
   Through our improvement of the gravitational and thermal tidal models, we can determine the dynamical state of exoplanets better, especially if they hold a thick atmosphere.
   In particular, the contribution of the atmospheric tides allows us to reproduce the spin state of Venus at a constant stellar luminosity.
   Our simulations have shown that the secular evolution of the spin and obliquity can lead to a retrograde spin of the Venus-like planet if the system starts from a high spin obliquity, in agreement with previous studies.
   The perturbing effect of a third body is still needed to determine the current state of Venus starting from a low initial obliquity.
   When the luminosity evolution of the Sun is taken into account, the picture changes.
   We find that the planet never reaches equilibrium: the timescale of the rotation evolution is longer than the luminosity variation timescale, which suggests that Venus may never reach a spin equilibrium state, but may still evolve.
   }


   \maketitle
%
\section{Introduction}\label{Sec:Introduction}

    The five thousand exoplanets discovered so far\footnote{\url{https://exoplanetarchive.ipac.caltech.edu/}} have revealed a great diversity of worlds. 
    As the number of discoveries continues to grow, an accurate modeling of exoplanets becomes increasingly important.
    In the context of the arrival of new powerful instruments such as the James Webb Space Telescope \citep[i.e. JWST;][]{Greene2016_JWST} and the Atmospheric Remote-sensing Infrared Exoplanet Large-survey mission \citep[i.e. ARIEL;][]{Tinetti_2021_ARIEL, Edwards_2022_ARIEL_target} in the characterization of rocky planets, we need to describe the dynamical state of rocky exoplanets with more realistic models by taking their internal structure and their potential atmosphere into account.
    A large number of the rocky planets discovered so far are in very close-in orbits, and are therefore usually considered to be in a circular and coplanar orbit and with a rotation that is synchronized with their mean motion, showing a permanent dayside.
    For planets with larger orbits, however, the rotational state and orbital elements (i.e., the  semi-major axis, eccentricity, orbital inclination, etc) evolve on a much longer timescale and are expected to have nonzero eccentricities and/or obliquities.
    Then, eccentricity or obliquity can trap the spin in a higher rotation state, that is in spin-orbit resonances (hereafter SORs), such as a 3:2 SOR, 2:1 SOR, or higher \citep[e.g.,][]{Makarov2013,Makarov2018}.
    If a planet has an atmosphere, another tidal mechanism must be taken into account: the atmospheric thermal tides.
    These are caused by the differential heating between day- and nightsides \citep{Gold_Soter_1969, ChapmanLindzen1970, DOBROVOLSKIS_1980, ingersoll1978, Correia_Laskar_2001, auclair_2017a}.
    This mechanism is a possible explanation for the current state of Venus as the thermal tides can both desynchronize the planet and increase its obliquity \citep[e.g.,][]{Correia_Laskar_2001, Correia_Laskar_2003}. 
    The rotation rate and the obliquity affect the climate of the planet by influencing the heat distribution.
    For example, spin rates faster than the synchronization can help prevent atmospheres from collapsing \citep[e.g.,][]{Wordsworth_2015} and change the fate of a potential surface water ocean \citep[i.e., complete vaporisation or not;][]{Turbet_2016} through a more effective heat redistribution in the atmosphere.
    We therefore need a complete dynamical framework with relevant tidal models to determine the rotation states of exoplanets as accurately as possible in the context of future data.
    
    In this article, we use the particular case of Venus to present our recent implementation of planetary tides \citep{Boue_Efroimky_2019} in a secular code called ESPEM (French acronym for Evolution of Planetary System and Magnetism, \citealt{benbakoura_2019,Ahuir_2021}).
    Here we study the case of a Venus-like planet around a Sun-like star.
    The rotation of Venus is thought to be an equilibrium between the gravitational bodily tides and thermal atmospheric tides \citep{Correia_Laskar_2001, correia_long-term_NSimulation_2003, correia_long-term_Theory_2003}.
    It is therefore a good laboratory for studying these mechanisms.
    Some unresolved issues still remain, however, such as whether the rotation of Venus is currently in equilibrium, and how it reached its current rotational state.
    The competition between the tides, gravitational and thermal, strongly depends on the internal state, but in the case of Venus, little is known about its internal structure.
    This will change with the next incoming mission to Venus, however, as EnVision \citep{EnVission_mission_2020}, DAVINCI \citep{Garvin_2022_DAVINCI}, and VERITAS \citep{VERITAS_mission} will bring valuable data about the internal state of Venus and the thermal atmospheric response of the planet \citep{Bills_2020}.
    
    To study the spin evolution of Venus-like planets, and in particular, the capture in SORs, it is necessary to describe the internal structure of the planet and atmosphere well.
    In particular, \cite{Walterova2020} showed that the internal structure also affects the SOR available by the planet.
    To ensure a good description in this work, we therefore computed the gravitational bodily tides using the formalism of \cite{Kaula_1964} along with the Andrade rheology \citep{Andrade1910}, using rheological parameters constrained from laboratory experiments on olivine \citep{Castillo-Rogez2011}.
    We computed the thermal tides using the analytical model of \cite{Leconte2015} adapted from the prescription developed by \cite{DOBROVOLSKIS_1980} to reproduce the current state of Venus.
    We also investigated the effect of the luminosity variation of the star on the equilibrium state between the gravitational and thermal tides.
    
    In Section~\ref{Sec:Method} we introduce the tidal model we used for the solid and thermal tides and the implementation in the ESPEM code.
    In Section~\ref{Sec:Solid_tides_Results} we discuss the evolution of the spin of a Venus-like planet when we only consider the influence of the solid tide.
    In particular, we discuss the well-known eccentricity-driven SORs and the less well-known inclination driven SORs.
    In Section~\ref{Sec:Atmopsheric_tides_Results} we discuss the evolution of the planet taking the thermal tides for the constant and evolving stellar luminosity into account.
    Finally, we discuss our findings and conclude in Section~\ref{Sec:discussion}.

\section{New model of planetary tides in ESPEM}\label{Sec:Method}

\begin{figure}[h!]%
        \centering
        \subfloat[\centering 
            Schematic representation of the atmospheric redistribution caused by the stellar heating on a synchronously rotating planet.
            The arrows show the movement of the particles of the atmosphere pushed from the hot spot (sub-stellar) toward the cold spots (morning and evening spots).]
            {\label{fig:scheme_thermal_tide}\includegraphics[width=0.4\textwidth] {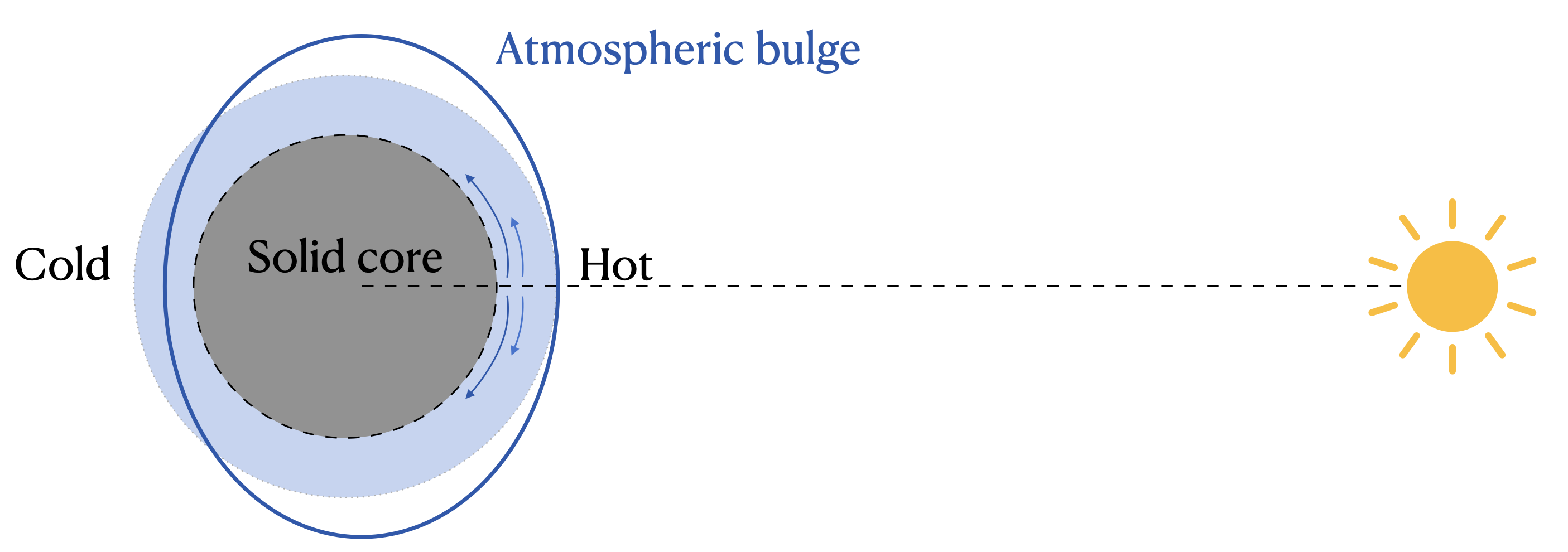}}%
        \qquad
        \subfloat[\centering 
            Delayed deformation of the atmosphere with respect to the position of the star (angle $\delta_a$ in the schema) due to the rotation of the planet. $\Omega$ is the spin rate of the planet, and is $n$ the mean motion and the star.] 
            {\label{fig:scheme_thermal_tide_rotate}\includegraphics[width=0.4\textwidth] {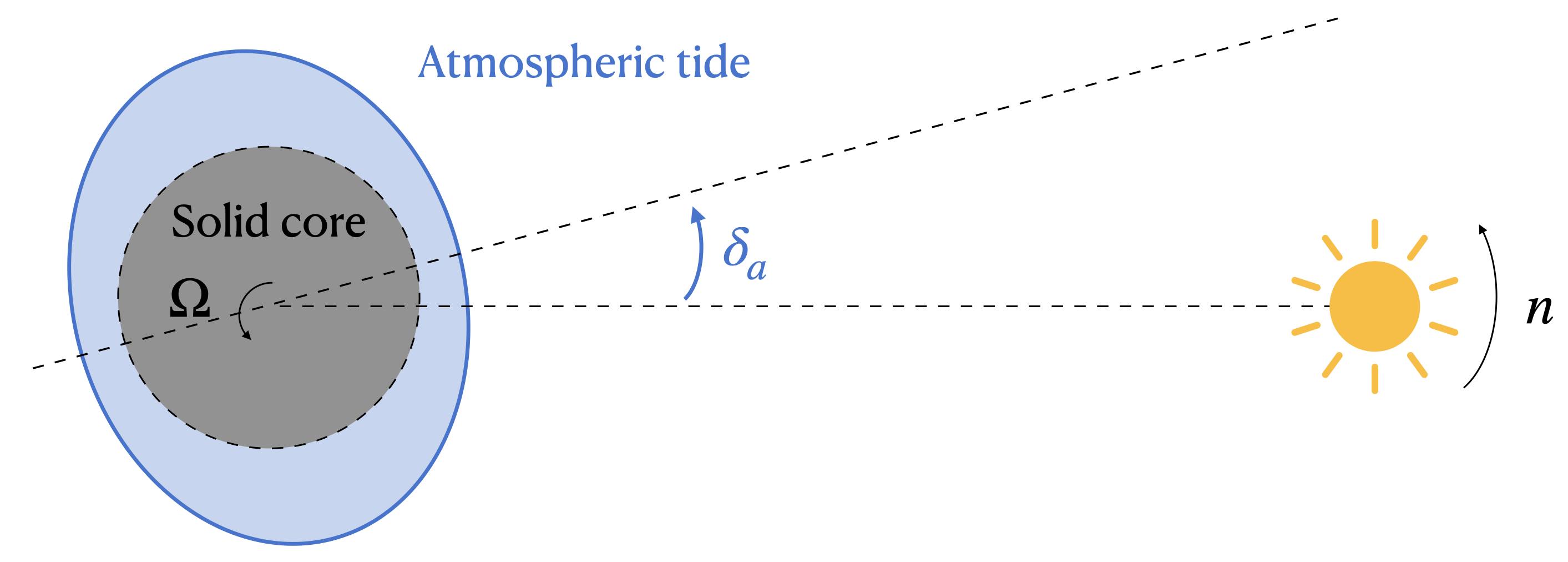}}%
        \qquad
        \subfloat[\centering
            Two tidal contributions, gravitational and thermal, acting in opposition on each other.]
            {\label{fig:scheme_thermal_tide_and_grav_tides}\includegraphics[width=0.4\textwidth] {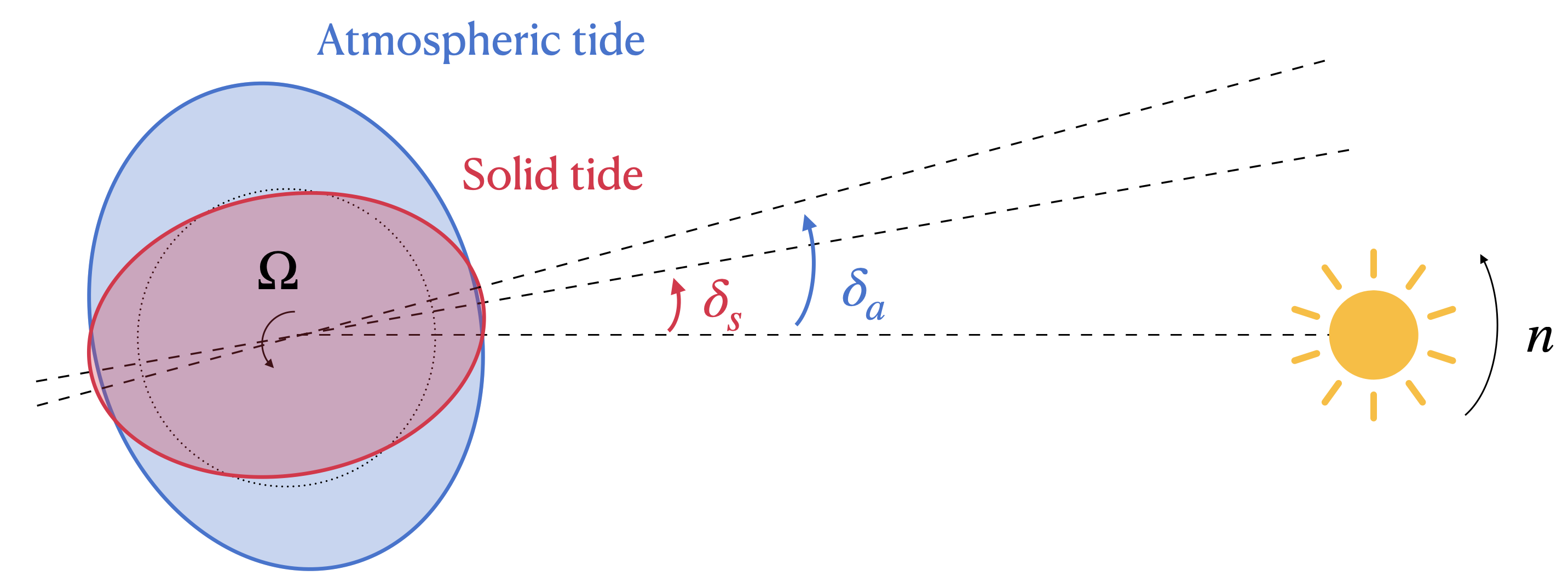}}%
        \caption{Tidal elongations of a rotating planet, composed of a solid core and a gaseous atmosphere, submitted to both gravitational and thermal forcing.
        Figures inspired from \cite{Correia_Laskar_2003}.}%
        \label{fig:atmospheric-grav-tides}
    \end{figure}

    We consider the equilibrium solid tides, which correspond to the mass redistribution of a body (i.e., the planet) under the influence of the gravitational perturbation of a massive (or close-in) orbiting body (i.e. the star).
    As the planet rotates, the solid bulge will be ahead from the position of the star (as illustrated in red in the Fig.~\ref{fig:scheme_thermal_tide_and_grav_tides}) if the spin of the planet is higher than the mean motion.
    Then, we consider the so-called thermal tides.
    These correspond to the mass redistribution of an atmosphere due to stellar heating.
    In the same manner in which the gradient of the gravitational potential causes the mass redistribution of the body, the thermal tides are raised by the differential heating through the atmosphere \citep{Gold_Soter_1969,DOBROVOLSKIS_1980,Correia_Laskar_2001}.
    The differential temperature between the day- and nightside causes a pressure gradient and therefore a mass redistribution of the atmosphere.
    This pressure gradient continuously redistributes the atmospheric particles from the high-temperature side (dayside) to the low-temperature side (nightside).
    As Fig.~\ref{fig:scheme_thermal_tide} shows, the direction of the bulge that forms is parallel to the direction of the heating source (i.e., the star).
    If the planetary spin is higher than the mean motion, as shown in Fig.~\ref{fig:scheme_thermal_tide_rotate}, the geometry of the deformation places the atmospheric bulge behind the position of the star by analogy with the solid deformation, while the solid bulge is ahead of the position of the star.
    The delayed response of the atmosphere caused by its radiative damping affects the dynamics of the planet through viscous coupling at the surface.
    Fig.~\ref{fig:scheme_thermal_tide_and_grav_tides} shows the combination of the gravitational and thermal tides on the planet.
    The two tides compete until an equilibrium is found.
    
    In the following, we introduce in Section~\ref{Subsec:Kaula_formalism} the formalism we used (Kaula) and in Section~\ref{Subsec:Tidal_love_num} the complex Love number, which allows us to express the tidal potential of the deformed planet in terms of Fourier series.
    In Section~\ref{Subsec:Atmospheric_tides_physics} we extend the notion of the potential Love number to a thermal Love number and detail the way in which we account for the influence of the thermal tides.
    In Section~\ref{Subsubsec:Hot_Homogeneous_profile} we design a homogeneous interior model for the planet that counterbalances the thermal tides for the excitation of Venus.
    In Section~\ref{Subsec:secular_equations} we discuss the corresponding orbital and rotational equations to finally described the implementation in the ESPEM code (Section~\ref{Subsec:Espem_code}).
    
 \subsection{Kaula formalism}\label{Subsec:Kaula_formalism}
\begin{figure}[H]
        \centering
        \includegraphics[width=0.5\textwidth]{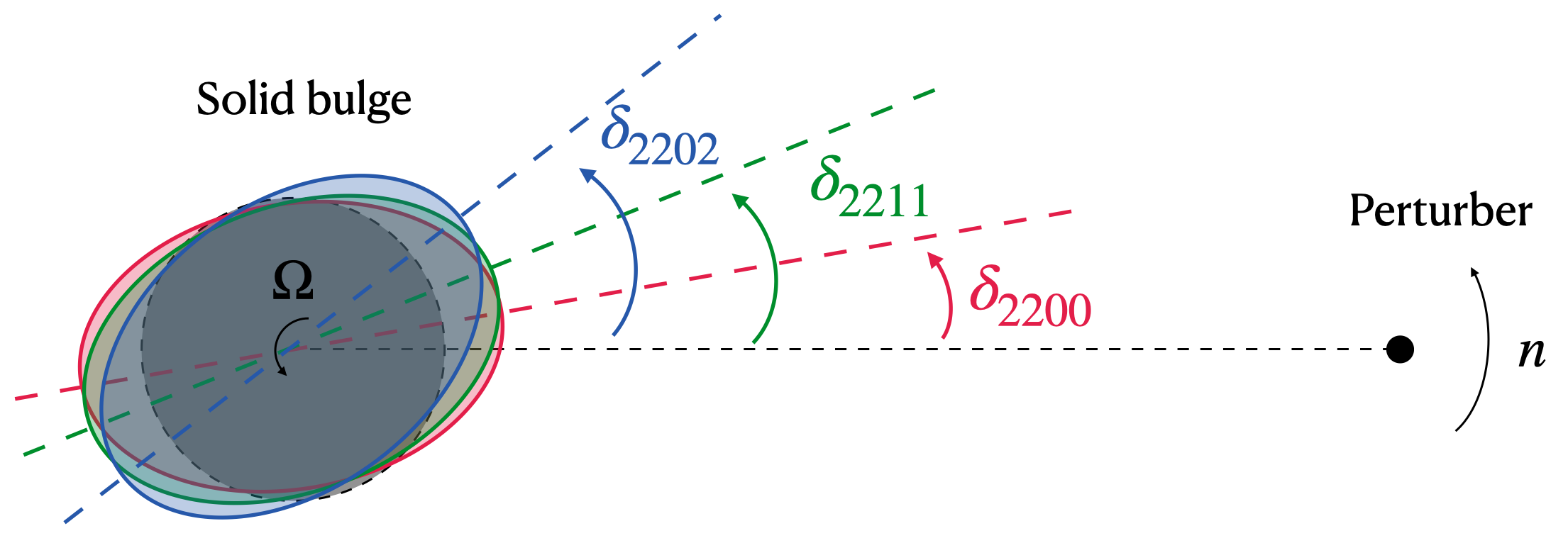}
        \caption{Schematic representation of the contribution of the tidal modes $l,m,p,q = (2200), (2211),~\text{and}~(2202)$ in the Kaula formalism.
        Each bulge represents the tidal deformation under a component of the tidal potential $U_{lmpq}$ of the perturber (point mass on the right.)
        }
        \label{fig:scheme_grav_tides_Kaula-frequ}
    \end{figure}

    In order to compute the tidal response of a body to a tidal perturbation, we need to use a formalism that is general enough to encapsulate the frequency-dependent response of a body.
    This response either requires a decomposition of the tidal potential created by the perturber (hereafter perturbing potential) into Fourier harmonic modes as developed by \cite{Kaula_1964}, or a time-domain approach as proposed by \cite{Correia2014} and \cite{Gervorgyan2020}.
    Both models allow a study of more complex and realistic rocky and icy bodies \citep{Efroimsky2013,Bolmont2020a}.
    We used the formalism developed by \cite{Darwin_1979} and adapted by \cite{Kaula_1964}, hereafter the Darwin-Kaula formalism.
    The Darwin-Kaula theory of bodily tides provides the expression of the perturbing potential of a disturbed body in Fourier series as 
    \begin{equation}
        \begin{split}
            U =& \sum_{l=2}^{\infty} \sum_{m=0}^l \sum_{p=0}^{l} \sum_{q\in \mathbb{Z}} U_{lmpq} (a, e, i, \sigma_{lmpq}), \\
        \end{split}
        \label{equ:Potential_fourier_modes}
    \end{equation}
    with a, e, and i, the semi-major axis, the eccentricity and the inclination respectively.
    The indexes $l,m,p,$ and $q$ are the indices of the harmonic modes, where $l, m$ are the orders associated with the associated Legendre polynomials, and $p,q$ the order of the Darwin-Kaula Fourier development.
    Each harmonic mode corresponds to an excitation frequency $\sigma_{lmpq}$, that is, the frequency with which the perturbing potential will affect the deformed body, defined as $\sigma_{lmpq} = (l-2p+q)n -m\Omega$ (with $\Omega$ and $n$ the spin rate and the mean motion respectively).
    Fig~\ref{fig:scheme_grav_tides_Kaula-frequ} shows the contribution of three different modes $(l,m,p,q)$, the $(2,2,0,0), (2,2,1,1),~\text{and}~(2,2,0,2)$ modes, which correspond to the frequencies $2(n-\Omega)$, $n-2\Omega$, and $4n-2\Omega$, respectively.
    The $(l,m,p,q) = (2,2,0,0)$ mode corresponds to the circular coplanar case (i.e., the semi-diurnal frequency). 
    The $(2,2,1,1)~\text{and}~(2,2,0,2)$ modes are two of the frequencies that are excited when the eccentricity is nonzero.
    
    This formulation is general and fundamental enough to be valid for an arbitrary rheology, and can also be used in the context of the thermal tides (see Sec.~\ref{Subsec:Atmospheric_tides_physics}).
    When the perturber is far enough away, we can keep the development of the gravitational potential at the quadrupolar order only, $l=2$ \citep{Makarov2013,Mathis_LePoncin-Lafitte2009}.
    We restricted the eccentricity expansion numerically up to the order 7, which corresponds to the index 7 in the summation over $q$ and eccentricities up to $0.3$.

 \subsection{Solid Love number}\label{Subsec:Tidal_love_num}
    The response of a planet to the tidal disturbance is quantified using the tidal Love number ${k_2}$ \citep{Love_1909}.
    The Love number links the perturbing potential and the additional potential created by the deformed planet in response to the perturbing potential.
    As we used the quadrupolar component of the tidal potential, we can write each mode as 
    \begin{equation}
        U_{2mpq}^{\text{tidal bulge}}(\sigma_{2mpq}) = \bar{k_{2}}(\sigma_{2mpq}) ~ U_{2mqp}^{\text{tidal perturbation}}(\sigma_{2mpq}),
        \label{eq:k2_link_potential}
    \end{equation}
    where $U_{2mpq}^{\text{tidal bulge}}(\sigma_{2mpq})$ corresponds to the quadrupolar component of the potential Eq.~\ref{equ:Potential_fourier_modes} ($l=2$), and $\bar{k_{2}}(\sigma_{2mpq})$ (hereafter $\bar{k_{2}}$) is related to the amplitude of the complex quadrupolar Love number \citep{Efroimsky2012a} at the quadrupolar mode $l=2$ with a frequency dependence with $\sigma_{2mpq}$.
    
    Physically, the quadrupolar Love number quantifies the response of a body submitted to a periodical external perturbation of frequency $\sigma_{2mpq}$.
    Here the periodical perturbation  corresponds to the tidal potential as described by Eq.\ref{equ:Potential_fourier_modes}.
    It can be written as a complex number, where the real part represents the pure elastic behavior and the imaginary part is the viscous behavior, written as
    \begin{equation}
        \bar{k_2} = \Re~(\bar{k_2}) + i~\Im~(\bar{k_2}) ~ = |\bar{k_2}|~\text{exp}\big(-i \epsilon_2 \big).
        \label{eq:Complex_Love_number}
    \end{equation}
    Thus, we can link the phase of the exponent $\epsilon_2$ with the angle between the tidal bulge and the position of the perturber with $\delta = \epsilon_2 /2$ \citep{RemusMathis_2012_Anelastic-dissip}.
    As shown in Fig~\ref{fig:scheme_grav_tides_Kaula-frequ}, each phase is associated with an excitation mode $(2mpq)$.
    
    This complex Love number can be computed for any density, shear modulus and viscosity profiles by integrating the equations of motions and Poisson's equation relating the displacement, stress, strain and induced potential in the frequency domain assuming a compressible Andrade rheology following the method described in \cite{Dumoulin2017} and \cite{Tobie2019}. 
    For a homogeneous solid body, the Love number can be determined from analytical solutions following \cite{Efroimsky2012b}.

    The tidal torque directly depends on the imaginary part of the Love number.
    In the circular coplanar case, the tidal torque applied on the planet is expressed as \citep{Kaula_1964,Goldreich1966,Murray_1999solar}
    \begin{equation}
        T^{\text{grav}} = \frac{3}{2}\frac{\mathcal{G} M_\star R_p^5}{a^6} \Im (\bar{k}_2),
        \label{eq:solid_torque_circular_coplanar}
    \end{equation}
    with $\mathcal{G}$ the gravitational constant, $M_\star$ the stellar mass, $R_p$ the planetary radius, $a$ the semi-major axis and $\Im (\bar{k}_2)$ the imaginary part of the Love number, which can be linked with the well-known dissipation factor $Q_2$ and the Love number modulus $k_2$ with $\Im (\bar{k}_2(\sigma)) = -k_2(\sigma)/Q_2(\sigma) \text{Sign}(\sigma)$ \citep{Gold_Soter_1966, Ogilvie_2014,Bagheri_2022}.

\begin{figure}[H]
        \centering
        \includegraphics[width=0.15\textwidth,angle=-90]{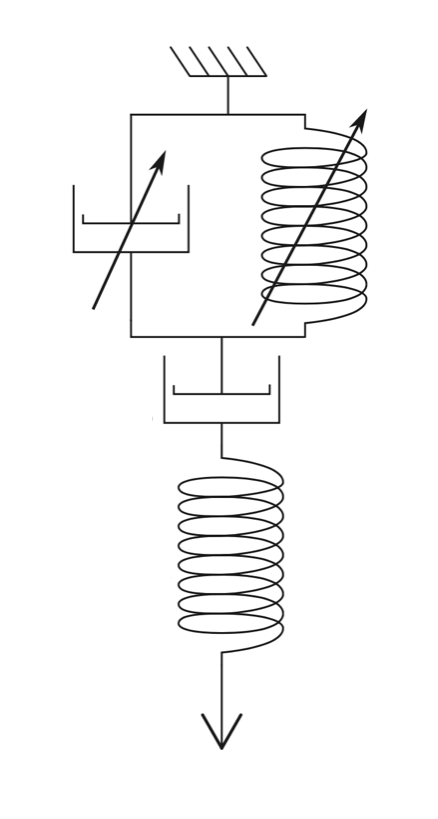}
        \caption{Schematic representation of the Andrade anelastic model used in this study \citep[adapted from][]{Renaud2018}. 
        The two first components in series represent a spring and a dashpot.
        The elements in parallel represent an infinite number of springs and dashpots.}
        \label{fig:Andrade_model}
    \end{figure}

    Then, we need to model an appropriate Love number $\bar{k_2}$ for rocky bodies in order to compute the secular tidal effects.
    Most models assume that planets are made of weakly viscous fluid \citep[e.g.,][]{Hut1981, Goldreich1966} even for rocky planets. 
    However, it has been shown that they do not reproduce the correct behavior for highly viscous solid bodies, such as the evolution of their rotation \citep{Henning2009,Efroimsky2013}.
    We used a more realistic rheological response, the Andrade rheology \citep{Andrade1910}, to better reproduce the behavior of a rocky body under periodical forcing \citep{Castillo-Rogez2011}.
    The Andrade rheology is an anelastic model built as a combination of dashpots and springs.
    It is composed with two first components in series, a dashpot and a spring which model the pure viscous damping and the pure elastic rigidity respectively, which correspond to the so-called Maxwell rheology \citep[e.g.,][]{Correia2014}.
    The Maxwell components are linked in series with an infinite number of springs and dashpots in parallel which correspond to the hereditary Andrade property, which retains some aspect of material memory (see Fig~\ref{fig:Andrade_model} and \citealt{Efroimsky2012a} for details).
    This model successfully reproduces a broad range of laboratory measurement of solid behavior under stress and strain, including silicate minerals, metals, and ices \citep{Andrade1910,Andrade1914,McCarthy_Castillo-Rogez2013}.
    The rheological profile used in this study was computed with a multilayer model following the method published by \cite{Tobie_2005, Tobie2019} and \citet{Bolmont2020b}.
    
    The Love number can be computed with the method of \cite{Bolmont2020a}.
    Following \cite{Efroimsky2012b}, the complex tidal Love number $\Bar{k}^{\text{grav}}_2$ is given by 
    \begin{equation}
        \Bar{k}^{\text{grav}}_2 = \frac{3}{2} \frac{1}{1+A_2 J/\Bar{J}} ~,
        \label{eq:complex_LoveNumber}
    \end{equation}
    with $J = 1/\mu$ the unrelaxed compliance (with $\mu$ the unrelaxed elastic shear modulus in Pa) and $A_2$ defined by  
    \begin{equation}
        A_2 = \frac{57 J^{-1}}{8 \pi \mathcal{G} \rho^2 R_p^2} ~,
        \label{eq:factor_A2}
    \end{equation}
    with $\rho$ the density.
    $\Bar{J}$ is the complex compliance of the material and defined in the formalism of the Andrade rheology with \citep{Castillo-Rogez2011}
    \begin{equation}
        \bar J  = J +\beta(i \sigma)^{-\alpha}\Gamma(1+\alpha) -\frac{i}{\eta \sigma} ~,
        \label{eq:andrade_complex_compliance}
    \end{equation}
    with $\beta$ a factor that describes the intensity of anelastic friction in the material, $\Gamma$ the Gamma function, $\eta$ the shear viscosity, $\sigma$ the excitation frequency, and $\alpha$ an experimentally fit parameter that which represents the frequency dependence of the transient response.
    A value of $\alpha$ in the range of $0.23$-$0.28$ allows us to reproduce the dissipation factor and $k_2$ for the Earth at different frequencies \citep{Tobie2019}.
    
    We studied the case of a Venus-like planet with different end-member temperature profiles.
    Little is known about the interior structure of Venus.
    This will likely improve with the upcoming ES EnVision mission \citep{EnVission_mission_2020} and the NASA DaVinci \citep{Garvin_2022_DAVINCI} and VERITAS \citep{VERITAS_mission} missions.
    In the meantime, we considered four possible structures.
    We used one multilayer profile (referred to as the reference profile) with Earth-like viscosity values as a reference, two other profiles with viscosity values divided by 10 or multiplied by 100 relative to the reference profile and one homogeneous profile (see Sec.~\ref{Subsubsec:Hot_Homogeneous_profile}).
    The multilayer structures can be considered as end members of what we think could be the real interior of Venus \citep[e.g.,][]{Bolmont2020a}.
    The Love numbers associated with the homogeneous profile where computed following the formula described in \cite{Bolmont2020a} and \cite{Efroimsky2012b} (see section \ref{Subsubsec:Hot_Homogeneous_profile}). 
    For the multilayer reference structure, we derived the radial density and seismic velocities in the mantle of the planet by using the Perple\_X code\footnote{\url{http://www.perplex.ethz.ch}} \citep{Connolly2005}, which uses a temperature profile from \cite{Armann2012} together with the shear modulus profile from the compositional model V1 of \cite{Dumoulin2017}.
    The viscosity was computed as a function of the temperature and pressure profiles as \citep{Dumoulin2017}
    \begin{equation}
        \eta = \frac{1}{2} A_0^{-1} d^{2.5} \mathrm{exp}\Big(\frac{E_a+PV_a}{RT}\Big),
        \label{eq:viscosity_formulae}
    \end{equation}
    with $E_a$ and $V_a$ the activation and volume energy, respectively, and $A_0$ the pre-exponential factor, which are parameters from the Arrhenius equation and depend on the material and $d$ the grain size.
    The parameters of the dry olivine considered in the upper mantle are $E_a = 300~\mathrm{kJ}~\mathrm{mol}^{-1}$, $A_0=6.08\times10^{-19}~\mathrm{Pa}^{-1}~\mathrm{s}^{-1}$, with a grain size $d=0.68~\mathrm{mm}$.
    
\begin{figure}[H]
        \centering
        \includegraphics[width=0.5\textwidth,trim = 1.5cm 0cm 2.2cm 1cm, clip]{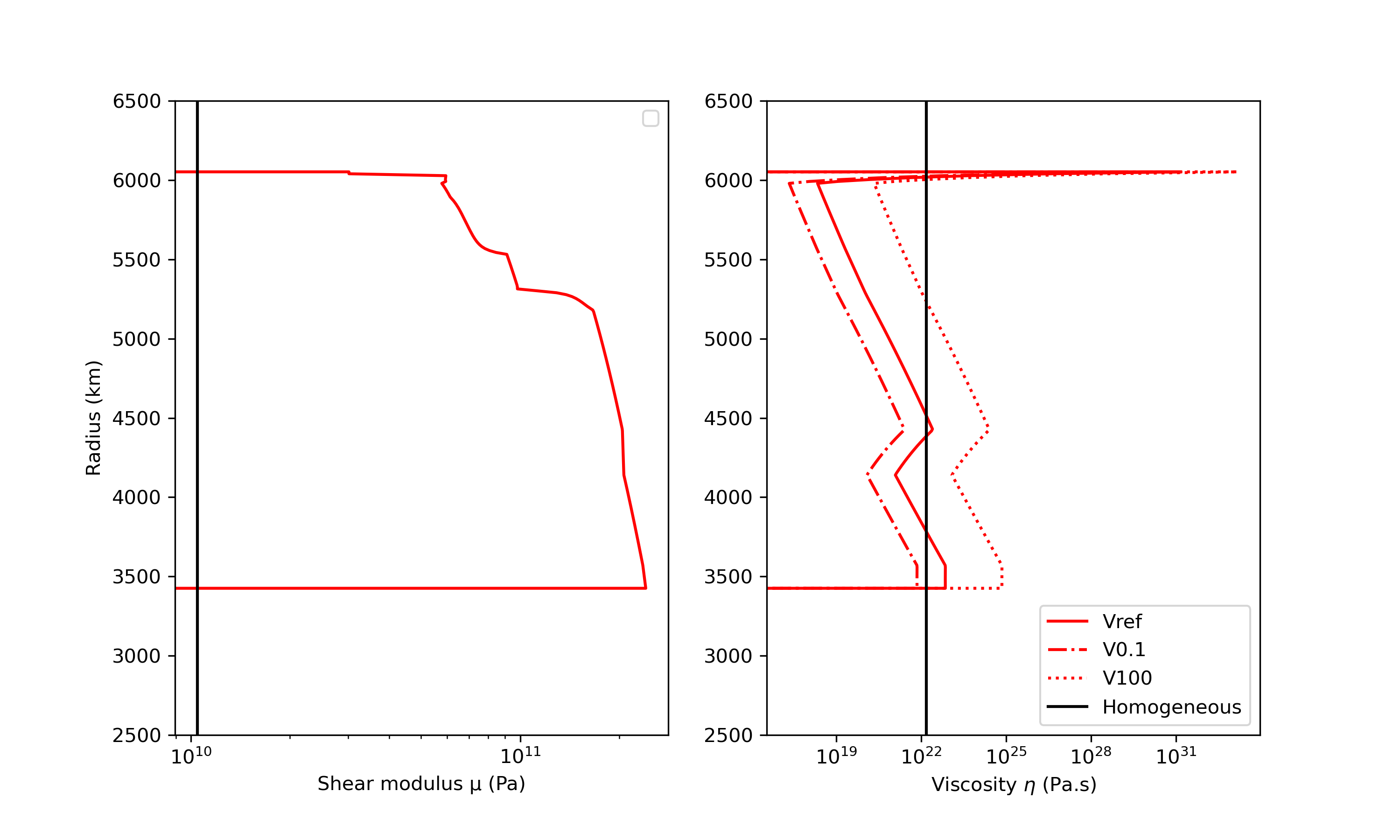}
        \caption{
        Shear modulus and viscosity profiles for the multilayer reference model Vref and homogeneous model considered here, shown as solid red and black lines, respectively.
        The  dotted and dash-dotted red lines represent the two profiles V0.1 and V100 derived from the multilayer reference model with a viscosity multiplied by x0.1 and x100 times, respectively.
        The viscosity $\eta$ is computed as in \cite{Dumoulin2017} using Eq.~\ref{eq:viscosity_formulae} of this work.
        }
        \label{fig:internal_cold-hot_profile}
    \end{figure}

\begin{figure}[H]
        \centering
        \includegraphics[width=0.45\textwidth,trim = 0.8cm 1cm 1.8cm 1cm, clip]{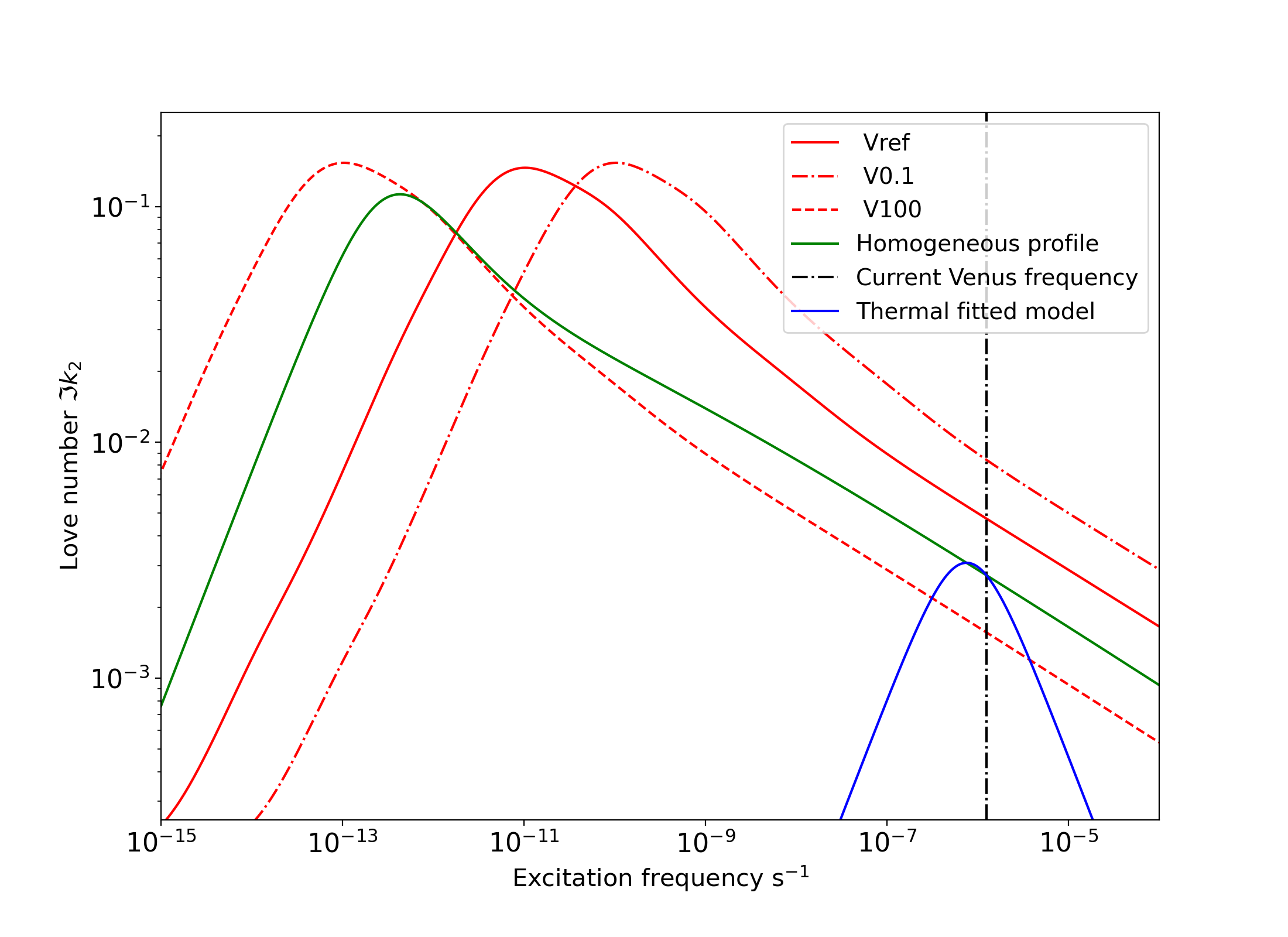}
        \caption{
        Imaginary part of the gravitational Love number $\Im(k_2^{\text{grav}})$ as a function of the excitation frequency of a Venus-like planet for different viscosity profiles.
        The multilayer reference profile Vref derived from \cite{Armann2012} is shown as the solid red line.
        The dotted and dash-dotted red lines represent the V0.1 and V100 profiles derived from the multilayer reference one presented in Fig~\ref{fig:internal_cold-hot_profile}.
        In blue we present the imaginary Love number $\Im(k_2^{\text{thermal}})$ as a function of the excitation frequency computed with Eq.~\ref{eq:imk2_convert_imqa} (in absolute values) associated with the amplitude of the thermal tides presented in Fig.~\ref{fig:Module_Lec_Atm}.
        The green curve represents the homogeneous profile described in Sec~\ref{Subsubsec:Hot_Homogeneous_profile}.
        The vertical dotted black line represents the absolute value of the current frequency state of Venus.
        }
        \label{fig:imk2_cold-hot_profile}
    \end{figure}

    Figure~\ref{fig:internal_cold-hot_profile} shows the internal profiles of the shear modulus $\mu$ and the viscosity $\eta$ for the homogeneous model (see Sec~\ref{Subsubsec:Hot_Homogeneous_profile}), the reference model (hereafter Vref) from \cite{Armann2012}, and two other models with viscosity profiles obtained by multiplying the viscous reference structure by 0.1 or 100 (denoted V0.1 and V100, respectively). 
    The metallic core structure was computed using PREM scaled to the Venusian pressure conditions \citep{Dumoulin2017}.
    The imaginary part of the Love numbers associated with these profiles were computed following \cite{Dumoulin2017}and \cite{Bolmont2020a} and are represented in Fig.~\ref{fig:imk2_cold-hot_profile}.
    A less viscous profile (dash-dotted line Fig.~\ref{fig:internal_cold-hot_profile}) that might correspond to a hotter mantle, will be more dissipative than a more viscous profile (dotted line Fig.~\ref{fig:internal_cold-hot_profile}) for frequencies higher than $10^{-11}$s$^{-1}$ (see Fig.~\ref{fig:internal_cold-hot_profile}).
    
  \subsection{Thermal Love number}\label{Subsec:Atmospheric_tides_physics}

    We considered only the feedback of the tidal bulge of the atmosphere deformed by the pressure gradient at the surface, considering that the atmosphere is perfectly coupled with the surface by viscous friction \citep[e.g.,][]{Leconte2015, Auclair_2019}.
    Thus, we neglected other feedbacks such as the effect of the pressure gradient on the shape of the solid crust and the gravitational anomaly of the atmosphere (see \citealt{Correia_Laskar_2003} for details).
    Because the mass redistribution of the atmosphere comes from the surface pressure anomaly, the imaginary part of the complex moment of the surface pressure field $\Im \big( \delta p_{s}^{2}\big)$ can be used as a prescription for the imaginary part of the thermal Love number \citep{Leconte2015, auclair_2017b}.
    The complex moment of the surface pressure field $ \delta p_{s}^{2}$ describes the thermal tides amplitude, and therefore, $\Im \big( \delta p_{s}^{2}\big)$ can be used to describe the dissipation.
    Then, we can relate $\Im \big(\delta p_{s}^{2}\big)$ to an imaginary thermal Love number $\Im \big(k^{\mathrm{thermal}}_2\big)$.
    The relation between these two quantities is discussed below.
    
\begin{figure}[H]
        \centering
        \includegraphics[width=0.46\textwidth,trim = 0.8cm 1cm 1.8cm 1cm, clip]{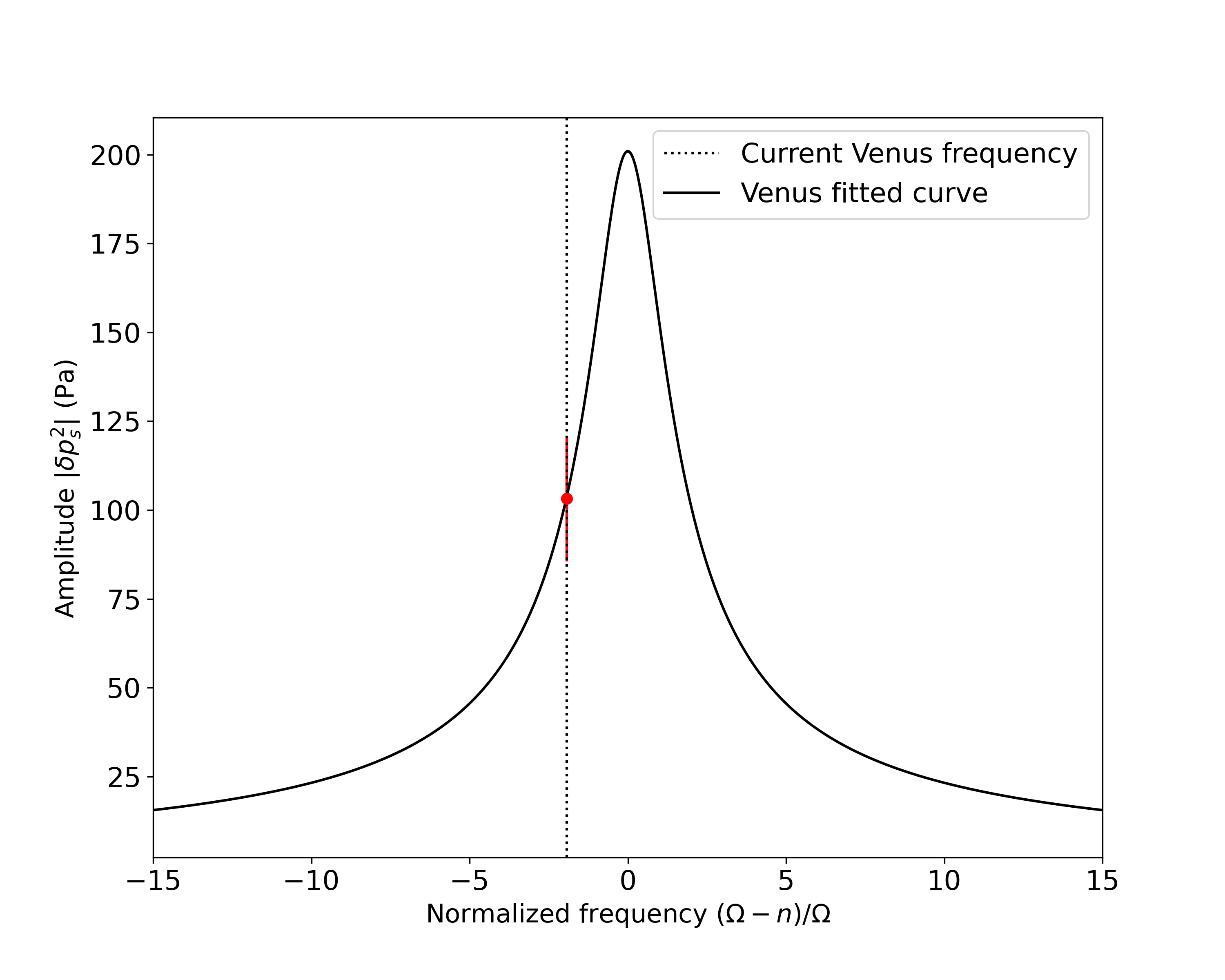}
        \caption{
        Amplitude of the pressure bulge $|\delta{p_{s}^{2}}|$ as a function of the normalized forcing frequency $(\Omega -n)/ \Omega$.
        The solid line represents the analytical solution fit for the point of Venus (red dot) computed with Venus GCM simulations \citep[see][]{Leconte2015}.
        The red bars on the Venus point are not strictly error bars.
        They represent the dispersion of the pressure bulge at the surface.
        }
        \label{fig:Module_Lec_Atm}
    \end{figure}

    To calculate the complex moment of the surface pressure field $ \delta p_{s}^{2}$, we used the work of \cite{Leconte2015}, who assumed a Maxwell-like frequency dependence \citep{ingersoll1978, Gold_Soter_1969, auclair_2017b}.
    More realistic frequency dependences have been proposed by \cite{Auclair_2019} as a generic formulation and a scaling law, adapted for $N_2$ atmospheres for different surface pressures.
    These models will be studied in future developments.
    \cite{Leconte2015} used a 3D climate model \citep[e.g.,][]{Leconte_2013_3DClimateModeling,Forget_2013_3DModelingEarlyMartian,Leconte_2013_RunawayGreenhouse} specifically tuned for the case of Venus to reproduce the amplitude of the thermal tides on Venus today and used this point to fit an analytical Maxwell-like solution.
    The analytical formulation of the module of the complex moment of the pressure field is expressed as 
    \begin{equation}
            \delta{p_{s}^{2}}  = -\frac{q_0}{1+i\frac{\sigma}{2\omega_0}},
        \label{eq:pressure_bulge_amplitude}
    \end{equation}
    such that the imaginary part can be written as 
    \begin{equation}
            \Im(\delta{p_{s}^{2}}) = \frac{q_0\frac{\sigma}{2\omega_0}}{1+\left(\frac{\sigma}{2\omega_0}\right)^{2}},
        \label{eq:imaginary_pressure_bulge_amplitude}
    \end{equation}
    with $q_0$ the amplitude of the quadrupole term of the pressure field at zero frequency, $\omega_0$ the radiative frequency, and $\sigma$ the excitation frequency.
    The radiative frequency can be identified with the inverse of the thermal equilibrium timescale.
    The parameters fit on the GCM simulation of Venus are: $q_0 = 201$~Pa and $\omega_0 = 3.77\times10^{-7}$~s$^{-1}$ \citep{Leconte2015}.
    Figure~\ref{fig:Module_Lec_Atm} shows the amplitude of the pressure bulge $|\delta{p_{s}^{2}}|$ as a function of the normalized forcing frequency.
    The solid curve shows the analytical solution fitted from the values of the amplitude and the phase lag computed with the GCM simulation of Venus of \cite{Leconte2015}.
    As it was not possible to run the specific GCM of Venus for different rotation states, it was not possible to constrain the Maxwell fit better.
    
    The thermal Love number $\Im(k_2^{\text{thermal}})$ can be determined from the complex moment of the surface pressure field $\Im \big( \delta p_{s}^{2}\big)$ by identification between the thermal and the gravitational torque.
    The expression of the thermal torque raised by the mass redistribution of the atmosphere is \citep[e.g.,][]{Gold_Soter_1966,Correia_Laskar_2001,Correia_Laskar_2003,Leconte2015,auclair_2017a}
    \begin{equation}
        T^{\text{thermal}} = \sqrt{\frac{24 \pi}{5}}\frac{M_{\star}}{M_{p}}\frac{R_p^6}{a^3} ~\Im\big(\delta p_{s}^{2}\big) ~,
        \label{eq:torque_atmospheric}
    \end{equation}
    with $M_\star$ the stellar mass, $M_p$ and $R_p$ the planetary mass and radius respectively, and $a$ the semi-major axis.
    The thermal equivalent Love number $\Im(k_2)$ can be written by identification with the expression of the solid torque (Eq.~\ref{eq:solid_torque_circular_coplanar}) as
    \begin{equation}
        \Im(k_2^{\text{thermal}}) = - \sqrt{\frac{32 \pi}{15}}\frac{a^3 R_p}{\mathcal{G} M_{\star}M_{p}}\Im\big( \delta p_{s}^{2}\big) ~,
        \label{eq:imk2_convert_imqa}
    \end{equation}
    We note that in contrast to $\Im(k_2^{\text{grav}})$, the thermal Love number $\Im(k_2^{\text{thermal}})$ depends on the semi-major axis of the planet as the intrinsic response of the atmosphere depends on the flux received by the planet.
    Then, the two torques, gravitational and thermal, do not have the same dependence as to the semi-major axis, which allows an equilibrium point at which the two tides compensate for each other.
    The dependence of the mass of the atmosphere is contained in the surface pressure term.
    We highlight that a more massive atmosphere does not necessarily lead to stronger atmospheric tides.
    For a more massive atmosphere, the atmospheric layers are more opaque to the stellar flux.
    Thus, as less stellar flux reaches the surface, the thermal tides are damped for a more massive atmosphere.
    This effect strongly depends on the composition of the atmosphere and requires a better model to be taken into account.
    A more massive atmosphere is not investigated in this study.
    The imaginary Love number $\Im(k_2^{\text{thermal}})$ as a function of the excitation frequency for the fitted analytical model (Eq.~\ref{eq:imk2_convert_imqa}) is plotted in blue in Fig.~\ref{fig:imk2_cold-hot_profile} in absolute values.
    In the following, we study the presence equilibrium points as a function of tidal frequency for different internal profiles.

    \subsection{Equilibrium state between gravitational and thermal tides}\label{Subsubsec:Hot_Homogeneous_profile}
    
\begin{figure}[h!]
        \centering
        \includegraphics[width=0.45\textwidth, trim = 0.3cm 0.2cm 0cm 0.2cm, clip]{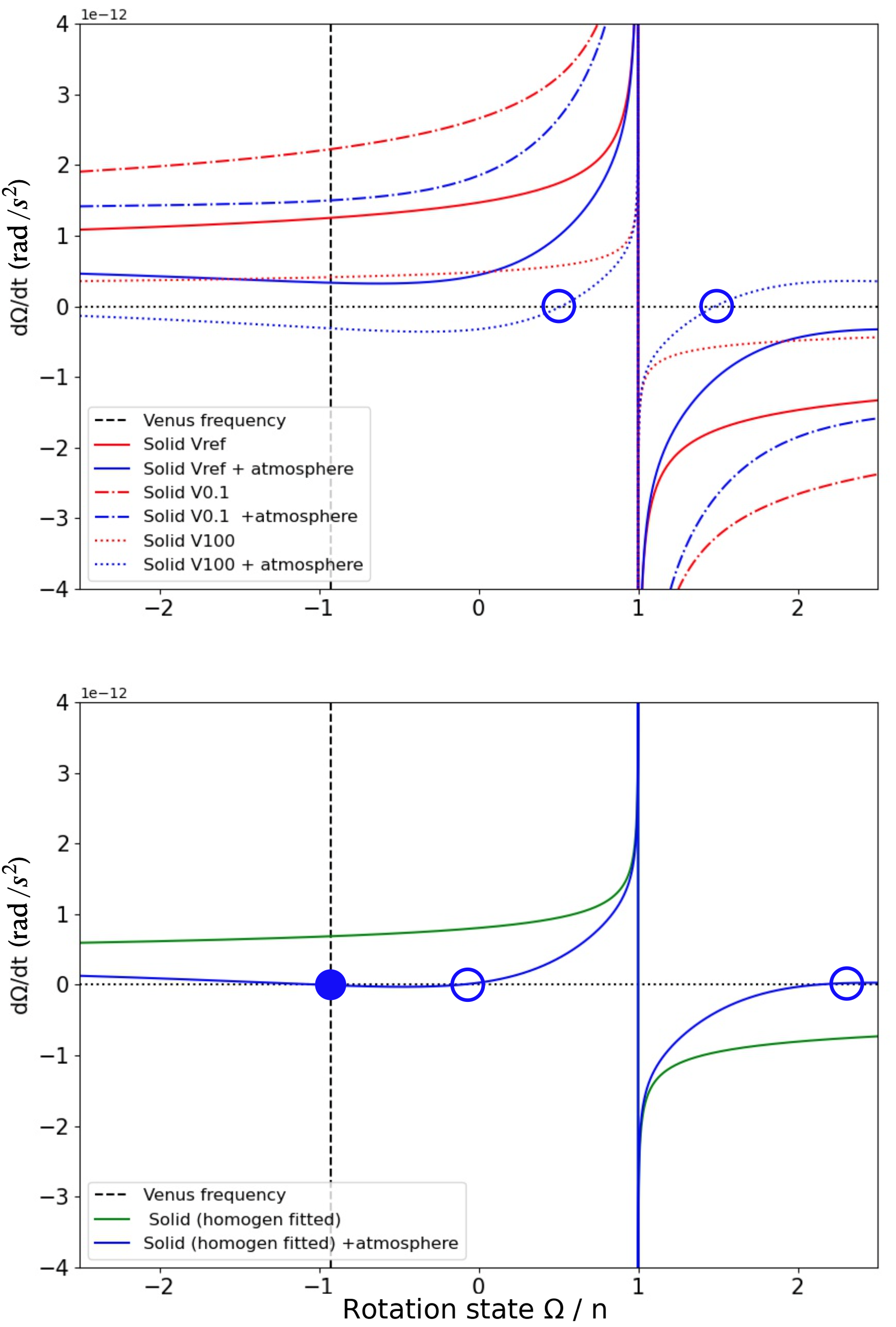}
        \caption{
        Spin derivative as a function of the rotation (in terms of $\Omega/n$, $\Omega$ and $n$ the planetary spin and mean motion respectively).
        In the top panel, the red lines correspond to the solid tides, and the blue lines correspond to the cases with solid and atmospheric tides.
        The solid lines correspond to the reference multilayer profile Vref. 
        The dotted and dash-dotted lines correspond to the V0.1 and V100 profiles, respectively.
        In the bottom panel, the green line represents the solid tides associated with the homogeneous body (see Sec.~\ref{Subsec:Tidal_love_num} for details).
        The blue line corresponds to the cases with the contribution of atmospheric tides.
        The vertical dashed black line represents the current frequency of Venus in both panels.
        The dots (filled and empty) represent the equilibrium states between the gravitational and thermal tides (stable and unstable, respectively).
        }
        \label{fig:Dspin_V1V4Hom_Lecfit}
    \end{figure}

    An equilibrium state between the gravitational and thermal tides can be determined by comparing their imaginary Love numbers.
    The two tides compensate for each other when the addition of the two imaginary Love numbers is $0$ (when the two absolute values are equal; see Fig.~\ref{fig:imk2_cold-hot_profile}).
    Figure~\ref{fig:imk2_cold-hot_profile} shows that the $\Im(k_2^{\text{grav}})$ corresponding to the multilayer reference profile (solid red curve) is always higher in amplitude than the $\Im(k_2^{\text{thermal}})$ (solid blue curve). 
    Figure~\ref{fig:Dspin_V1V4Hom_Lecfit} shows the spin derivative for the multilayer reference profile, the V0.1 and V100 profiles, and the homogeneous fit profile.
    Comparing the cases without and with atmosphere (in red and blue, respectively), we find that the solid tides corresponding to the multilayer reference profile are strong enough to compensate for the thermal tides at any frequency.
    On the one hand, the V0.1 profile is also sufficiently dissipative to compensate for the thermal tides, as it corresponds to a more dissipative interior and thus stronger solid tides.
    Thus, the spin derivatives associated with the reference profile and the less viscous one are not in equilibrium.
    The system will therefore evolve to the 1:1 SOR.
    On other the hand, the V100 profile leads to weaker solid tides.
    In this case, the spin derivative in Fig.~\ref{fig:Dspin_V1V4Hom_Lecfit} shows that the thermal tides are sufficient to compensate for the gravitational tides, except close to the synchronization, where the solid tides are still strong enough to make the 1:1~SOR stable.
    This profile, which might correspond to a colder mantle, is not dissipative enough to compensate for the thermal tides close to the current state of Venus.
    This is shown in Fig.~\ref{fig:imk2_cold-hot_profile}, where the $\Im(k_2^{\text{thermal}})$ (solid blue curve) has a broad range of frequencies (from $10^{-7}s^{-1}$ to $2\times10^{-5}\text{s}^{-1}$), where it dominates the gravitational Love number $\Im(k_2^{\text{grav}})$ associated with the V100 profile (dashed red curve).
    Thus, the two intersection points correspond to possible equilibrium states, at which the two tides compensate for each other.
    The equilibrium points shown in the top panel of Fig~\ref{fig:Dspin_V1V4Hom_Lecfit} (empty blue circles) correspond to the intersection point at low frequency (about $10^{-7}~$s$^{-1}$ on Fig~\ref{fig:imk2_cold-hot_profile}).
    Considering the slope of the derivative, however, this point is not stable.
    The second point at high frequency (about $4.5\times10^{-5}~$s$^{-1}$ on Fig~\ref{fig:imk2_cold-hot_profile}) corresponds to a fast rotation of about three days.
    This last point is not investigated further because, on the one hand, this state is far from the current state of Venus, and on the other hand, it belongs to the high-frequency regime.
    In our case, the Maxwell-like frequency-dependent model of thermal tides overestimates the strength of the thermal tides at high frequencies, as the model was fit for the low-frequency regime \citep{Auclair_2019}.
    We therefore consider our approach to be valid in the low-frequency range, that is, for $|\Omega/n| < 15$.
    As this equilibrium point is very far from the present-day rotation of Venus, it would require a more complex model that is beyond the scope of this study, and we did not investigate it further.
    A better model, such as the parameterized model proposed by \cite{Auclair_2019}, will be studied in the future.

    None of the profiles reproduce the balance between the two contributions, gravitational and thermal, close to the frequency of Venus
    Using the method of \cite{Bolmont2020a} described in Sec~\ref{Subsec:Kaula_formalism} (Eq.~\ref{eq:complex_LoveNumber} to~\ref{eq:andrade_complex_compliance}), we fit a homogeneous profile (in density, viscosity, and rigidity) that reproduces an equilibrium point at the Venus frequency.
    As the profile we tried to construct is relatively close to the multilayer profile, we used the parameters given in Table 2 of \citet{Bolmont2020a} for this profile of Venus at $\alpha=0.25$ and only fit the value of the viscosity parameter $\text{log}(\eta)$.
    The other parameters are the rigidity $\text{log}(\mu)=10.02$ in Pa and the ratio of the Andrade and Maxwell time $\tau_A /\tau_M=0.89$ \citep{Castillo-Rogez2011}. 
    The homogeneous profile that fits the thermal Love number $\Im(k_2^{\text{thermal}})$ at the Venus frequency is found with $\text{log}(\eta) = 22.18$ and is plotted in Fig.~\ref{fig:imk2_cold-hot_profile} (green curve). 
    Fig~\ref{fig:Dspin_V1V4Hom_Lecfit} (bottom panel) shows the stable equilibrium point between the gravitational tides associated with the fit profile and the thermal tides (filled blue dot) at the current frequency of Venus as well as two unstable points (empty 
    blue points).
    These points correspond to the rotation states in which the two tides compensate for each other.
    We must highlight that the homogeneous profile allows for spin equilibrium at the current frequency of Venus in a very narrow range of internal states.
    As the interior temperature profile evolves on geologic timescales, it will be relevant to take the associated change of dissipation due to progressive cooling into account, which evolves on timescales of 100~Myr \citep{Bower_2019}, or radiogenic decay, tidal heating, and so on, to fully characterize the spin equilibrium.
    This temperature dependence will be addressed in future studies.
    
 \subsection{Secular equations}\label{Subsec:secular_equations}
    The secular equations we implemented were derived from the Hamiltonian formalism by \cite{Boue_Efroimky_2019}.
    We used Eqs.~(116) to (123) of their work, which were derived within the Darwin-Kaula formalism (see Appendix \ref{appendix:secular-equation}).
    One important hypothesis, that was formulated to derive these equations is the gyroscopic approximation, which implies that the spin rate of a body is much faster than the evolution of the spin-axis orientation.
    This approximation invalidates the equations when the spin tends to zero for a noncoplanar orbit.
    A singularity occurs when the spin rate is zero within this approximation \citep[see][]{Boue_Efroimky_2019}.
    The validity of the equations for an inclined orbit close to the null rotation state will need to be revisited.

    In this formalism, the inclination is defined by the angle between the orbital plane and the equatorial plane, which in other words corresponds to the angle between the orbital angular momentum and the planetary spin angular momentum\footnote{This angle is also often referred to as the obliquity, for instance, the obliquity of the Earth is about 23 degrees. In this study, we would thus say that the inclination of the Earth is 23 degrees.}.
    
    The development of \cite{Boue_Efroimky_2019} also includes the deformation of the secondary under the tidal effect of the primary. 
    We neglected the tidal deformation of the secondary.
    The resulting secular equations of the spin, eccentricity, and orbital inclination are presented in Appendix \ref{appendix:secular-equation}.

 \subsection{Implementation in the ESPEM code}\label{Subsec:Espem_code}
    We implemented the secular equations of \cite{Boue_Efroimky_2019} in the code ESPEM \citep{benbakoura_2019,Ahuir_2021}.
    This is a secular code integrating the dynamical evolution of a star-planet system.
    The code takes the coupling between the two layers of low-mass stars into account (convective and radiative layers), as well as the effect of the stellar wind and the torque due to the tides raised by the planet on the convective envelope star and the torque due to the star-planet magnetic interactions for circular and coplanar orbits \citep{Ahuir_2021}.
    The code was only used to compute the angular momentum exchange between the planetary orbit and the stellar radiative core and convective envelope angular momentum \citep{benbakoura_2019,Ahuir_2021}.
    We have added the tidal torque of the star on the planet within the formalism described in this paper.
    In addition to the equation for the semi-major axis, we also implemented the equations governing further osculating elements of the planet, such as the eccentricity, orbital inclination, longitude of ascending node, argument of periapsis, and planetary spin.
    The equations for spin, eccentricity, and inclination, longitude of ascending node and argument of periastron can be found in the Appendix~\ref{appendix:secular-equation}, Eq.~\ref{appendix:equation-spin_a} to \ref{appendix:equation-incl_a}.

    The user needs to provide a data file describing the time evolution of the mass and radius of the star as well as the evolution of the mass and radius of the radiative and convective envelopes, of the moment of inertia and the stellar luminosity.
    The stellar evolution is computed with evolution files provided by the code STAREVOL \citep{Amard2016}, which gives the internal dissipation and the evolution of the stellar quantities (e.g., the mass, radius of the radiative core, and convective envelope, and luminosity).
    The evolution of the stellar luminosity is used in Sec~\ref{subsec:Atmopsheric_tides_LuminosityVariation}.
    The user also needs to specify the initial conditions of the osculating elements of the planet as well as the rotation rate.
    The code also needs a data file describing the frequency dependence of the real and imaginary parts of the tidal Love number of the planet.
    This latter file should also provide the mass, radius, and the radius of gyration of the planet (which represents the internal density distribution).
    The Love numbers provided in these data files were computed with the method described in Sec~\ref{Subsubsec:Hot_Homogeneous_profile}.
    As explained in Sec~\ref{Subsec:Kaula_formalism}, several frequencies are excited depending on the eccentricity and inclination. 
    These frequencies were computed for each time step.
    The real and imaginary parts of the Love number were interpolated linearly from their frequency dependence in the data file.
    These interpolated values were then used to compute the derivatives of all the quantities mentioned before.

    We used the parameters of a Sun-Venus system, that is, a Venus-like mass and radius planet orbiting at $0.723$~AU.
    The parameters we used are listed in table~\ref{tab:initial_parameters}.
    The initial spin rate is shown from $\Omega/n=2.1$ for the ESPEM simulations as the SORs we studied are below this spin rate.
    The longitude of the ascending node and argument of pericenter, were set to zero.
    
    \begin{table}[h!]
        \caption{Numerical values used in the case of a Sun-Venus-like system.}
        \label{tab:initial_parameters}
        \begin{tabular}{ll}
        \hline
        \hline
         Parameter & Values \\ 
        \hline
         Star Mass ($M_\odot$) & 1 \\
         Planet Mass ($M_{\text{Earth}}$) & $0.815$ \\
         Planet Radius ($R_{\text{Earth}}$) & $0.857$  \\
         Semi-major axis (AU) & $0.723$ \\
         Eccentricity & $\{0, 0.1, 0.2 \}$ \\
         Spin Inclination (degrees) & $\{0, 5, 50, 120, 130 \}$ \\
        \hline
        \end{tabular}
    \end{table}
    
    The effect of the stellar tides, stellar wind or the evolution of the stellar layers were not taken into account in this study.
    We focused here only on the evolution of the rotation state of the planet under the tidal perturbation of a star.
    We considered this approach appropriate for studying the evolution of the planetary system we consider here.
    The effect of the Venusian tides inside the Sun can be considered to be negligible as the corresponding evolution timescale is about $10^{15}$~Gyrs order of magnitude\citep[e.g.,][]{bolmont_mathis_evolution_stellar_tides_2016}.

\section{Impact of the gravitational tides alone}\label{Sec:Solid_tides_Results}

    First, we investigated the secular evolution of the spin, the eccentricity, and the spin inclination of a Venus-like planet orbiting a Sun-like star driven by the gravitational tides alone.
    In other words, we first neglected the influence of the thermal tides, which is equivalent to first considering an atmosphereless planet. 
    In Section~\ref{Subsec:Solid_tides_EccentricSOR_Low_inclination} we discuss spin-orbit resonances for coplanar eccentric orbits, and in Section~\ref{Subsec:Solid_tides_InclinedSOR_Low_inclination}, we discuss spin-orbit resonances for inclined circular orbits.
    
 \subsection{Eccentricity-driven spin-orbit resonances}\label{Subsec:Solid_tides_EccentricSOR_Low_inclination}
 
    \cite{Hut1981} showed that if the orbit is eccentric, the planet reaches a pseudo-synchronization state, where the rotation rate of the planet is comparable to the mean motion around the periastron.
    The use of a model more appropriate for a highly viscous object results in discrete stable spin states in presence of eccentricity, however, in particular, SORs \citep{Makarov2013}.
    The higher the eccentricity, the higher the SOR order.
    A planet beginning its evolution with a high eccentricity and a spin faster than 2.5 times its orbital motion first becomes trapped in the 5:2~SOR.
    Then, as the eccentricity diminishes, the planet leaves the resonance to be trapped in the lower resonance, the 2:1~SOR, then leaving this configuration for a lower SOR, the 3:2 SOR as the eccentricity continues to decrease.
    Then, as the eccentricity continues to decrease, the rotation eventually becomes trapped in the 1:1 SOR, also known as the synchronous state, or tidal locking \citep[see also ][with the Creep tidal model]{Gomes2021}.
    
\begin{figure}[H]
        \centering
        \includegraphics[width=0.475\textwidth,trim = 0.0cm 1.6cm 2.5cm 3.0cm, clip]{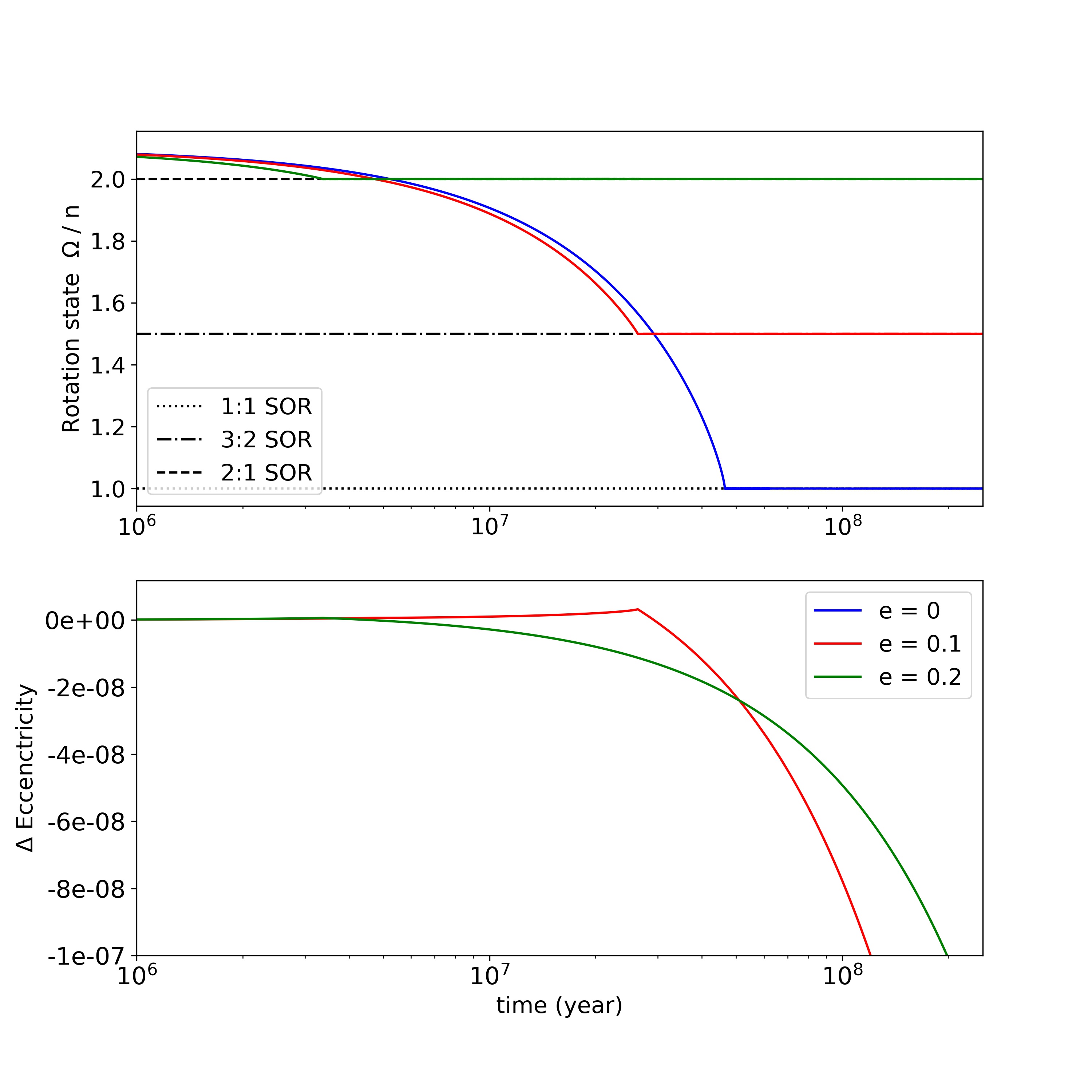}
        \caption{
            Evolution of a Venus-like planet with an initial rotation of $100$ days for three initial eccentricities (i.e., the null eccentricity, and the $0.1$ and $0.2$ eccentricities).
            The top panel shows the evolution of the planet rotation rate in terms of $\Omega /n$ ($\Omega$ and $n$ are the spin and mean motion, respectively).
            The bottom panel shows the variation in eccentricity $\Delta e$ (i.e., $(e -e(0))/e(0)$).
            }
        \label{fig:Espem_spin_ecc_evol}
    \end{figure}

    Figure~\ref{fig:Espem_spin_ecc_evol} shows the evolution of the rotation state and the eccentricity of a Venus-like planet with the multilayer internal reference structure (see Sec.~\ref{Subsec:Tidal_love_num}) for three initial eccentricities ($0.0$, $0.1$, and $0.2$) in coplanar orbit.
    The simulations started with an initial semi-major axis of $0.723$~AU and an initial rotation period of $100$ days.
    The figure shows that an eccentricity of $0.2$ is sufficient to allow the planet to be captured in the 2:1 SOR and the 3:2 SOR for an eccentricity of $0.1$.
    The planet can stay in this SOR as the eccentricity remains high enough throughout the simulation, as shown in the bottom panel of Fig.~\ref{fig:Espem_spin_ecc_evol}.
    
\begin{figure*}[h]%
        \centering
        \includegraphics[width=\textwidth,trim = 0.0cm 0.0cm 0.0cm 0.0cm] {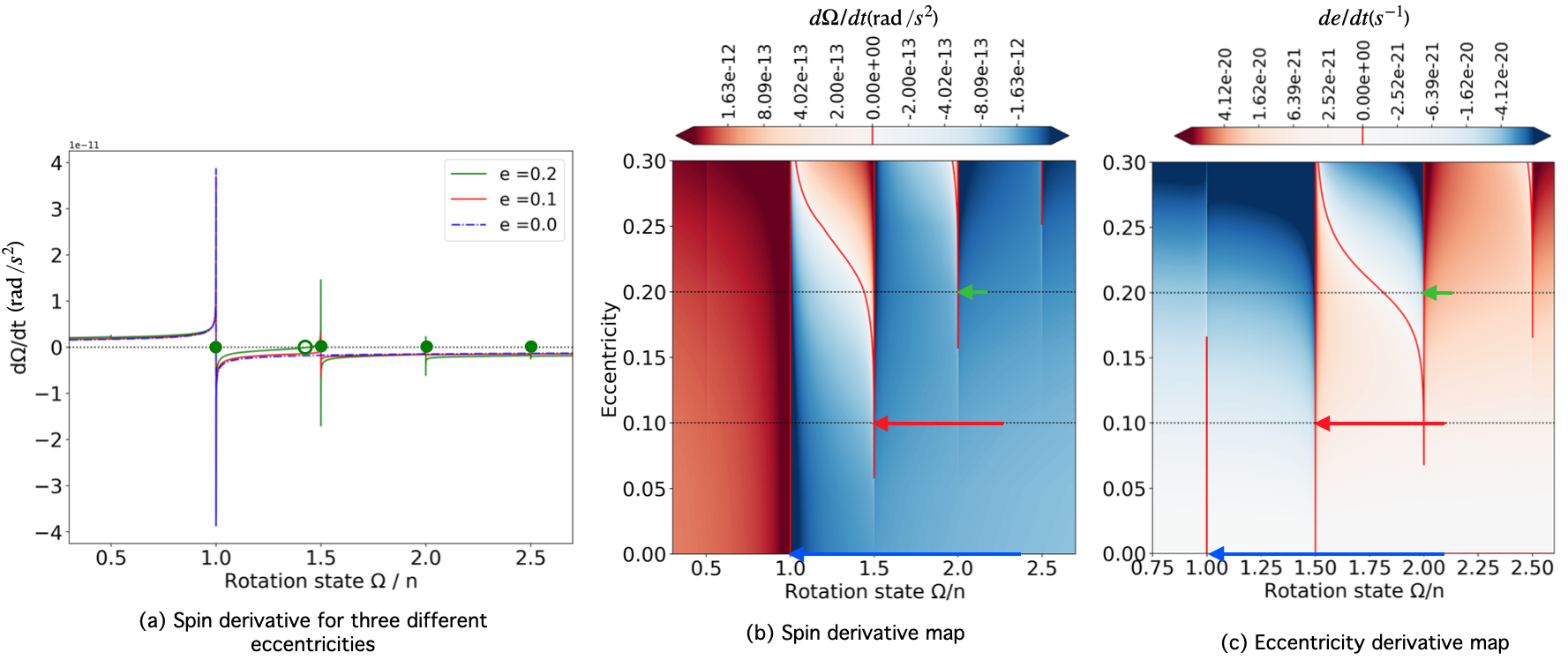}%
        \caption{
            Spin derivative $\mathrm{d}\Omega/\mathrm{d}t$ ($rad/s^2$).
            The left panel shows the spin derivative as a function of the rotation state $\Omega/n$ ($\Omega$ and $n$ are the spin and mean motion, respectively). for different eccentricities.
            The green dots (filled and empty) represent the equilibrium states, stable (i.e., SORs) and unstable, respectively.
            The middle and right panels represent the spin derivative and the eccentricity derivative as a function of the rotation state $\Omega/n$ and the eccentricity, respectively.
            The red colored areas depict the positives values, the blue areas depict the negative values, and the red line corresponds to $\mathrm{d}\Omega/\mathrm{d}t=0$.
            The dotted lines represent the two eccentric cases in the left panel ($e=0.1$ and $e=0.2$).
            The arrows represent the evolutions presented in Fig.~\ref{fig:Espem_spin_ecc_evol}.}%
        \label{fig:spin_ecc_3Panel}
    \end{figure*}

    The order of the resonance in which the planet capture is shown when we plot the spin derivative as a function of $\Omega/n$ ($\Omega$ and $n$ the planetary spin and mean motion respectively).
    Figure \ref{fig:spin_ecc_3Panel} shows how this quantity evolves with $\Omega/n$ for a fixed eccentricity (Fig.~\ref{fig:spin_ecc_3Panel}a) and how it evolves with $\Omega/n$ and for different eccentricities (Fig.~\ref{fig:spin_ecc_3Panel}b). 
    Figure~\ref{fig:spin_ecc_3Panel}a shows that for the circular case ($e=0$, dotted blue line), only one value of the spin result in $d\Omega/dt=0$, and therefore, only one possible equilibrium for the rotation state.
    The equilibrium is centered at $\Omega / n = 1$, which corresponds to the synchronization state.
    Higher eccentricities raise other resonances at a higher spin rate.
    For example, the $0.2$ eccentric case (in green on Fig.~\ref{fig:spin_ecc_3Panel}a) shows the 3:2 SOR and the 2:1 SOR in addition to the 1:1 SOR.
    The 5:2 SOR is also present, but the eccentricity must be higher than $0.25$ to keep this configuration stable, as the middle panel shows.

    Figure~\ref{fig:spin_ecc_3Panel}b shows the values taken by the spin derivative on a 2D map, as a function of $\Omega /n$ and the eccentricity, from $0.0$ to $0.3$.
    As we are restricted numerically up to the order $7$ in the eccentricity expansion, we computed the evolution up to $e=0.3$.
    This is sufficient because the population of rocky exoplanets does not present extreme eccentricities.
    The equilibrium points can be found with the null torque in red.
    The stable equilibrium states must satisfy the condition of a positive torque (in red) to its left and a negative torque (in blue) to its right.

    Figure~\ref{fig:spin_ecc_3Panel}b shows that increasing eccentricity allows higher-order SORs.
    The synchronization is accessible with $e=0$, while the 3:2 SOR becomes accessible at $e=0.06$, the 2:1 SOR at $0.16$, and the 5:2 SOR at $0.26$.
    The eccentric cases of the Fig.~\ref{fig:spin_ecc_3Panel}a are overplotted in the color map with the two horizontal dotted black lines.
    The evolutions shown in Fig. \ref{fig:Espem_spin_ecc_evol} are plotted in Fig.~\ref{fig:spin_ecc_3Panel}b and \ref{fig:spin_ecc_3Panel}c with the three colored arrows (with identical colors in the two figures).
    The spin quickly decreases in the SOR associated with its eccentricity.
    Because the eccentricity is damped by the tides, the SORs remain until the eccentricity becomes too low to stably maintain these configurations.
    Figure~\ref{fig:spin_ecc_3Panel}c represents the eccentricity derivative map in the eccentricity versus rotation state plane.
    In red we show the area in which the eccentricity increases, and in blue the area in which it decreases.
    In particular, the eccentricity appears to be slightly excited for the $e=0.1$ case shown in Fig.~\ref{fig:Espem_spin_ecc_evol}.
    This behavior can be explained with the derivative map of Fig.~\ref{fig:spin_ecc_3Panel}c.
    The eccentricity in the $e=0.2$ case of Fig.~\ref{fig:Espem_spin_ecc_evol} also appears to be excited before the state when the rotation reached the $2:1$~SOR.
    
    The timescale of the eccentricity evolution is too long.
    The departure from resonant states is therefore not shown here.

    We confirmed the eccentricity-driven spin-orbit resonances, such as the 1:1, 3:2, 2:1, and 5:2 SORs, and their dependence on the value of the eccentricity.
    Our results are also consistent with the work of \cite{Walterova2020}.
    We reproduce the eccentricity-driven resonances they showed for the shear modulus and viscosity of our fitted hot profile.
    \cite{Walterova2020} pointed out that the internal composition of the planet affects the stability of the SORs.
    Thus, the thermal evolution of the internal structure should be investigated, starting from a warm to a colder profile.
    As the temperature drives the viscosity and melt fraction of the mantle, the effect of the tidal heating should also be investigated and will be implemented in future developments.
    Then, the effect of the tidal heating should also be studied, but the tidal dissipation is not thought to be important for the case of Venus.
    Tidal dissipation is probably stronger for very close-in planets.
    The effect of the tidal heating of these planets will be the subject of future studies.
    
 \subsection{Inclination-driven spin-orbit resonances}\label{Subsec:Solid_tides_InclinedSOR_Low_inclination}
 
    The inclination-driven SORs has been discussed in \cite{Boue2016} in the context of gas giant planets responding to a Maxwell rheology.
    We show here that this behavior is also found for rocky planets with a rheology more adapted to rocky planets (Andrade), thus generalizing the findings of \cite{Boue2016}.
    This is the first study of inclination-driven SORs with a realistic rheology for rocky exoplanets that generalizes the first study of \cite{Boue2016} for giant planets, who used a simple Maxwell rheology.
    \begin{figure}[H]
       \centering
       \includegraphics[width=0.475\textwidth, trim = 1.3cm 1.8cm 2.5cm 3.0cm, clip]{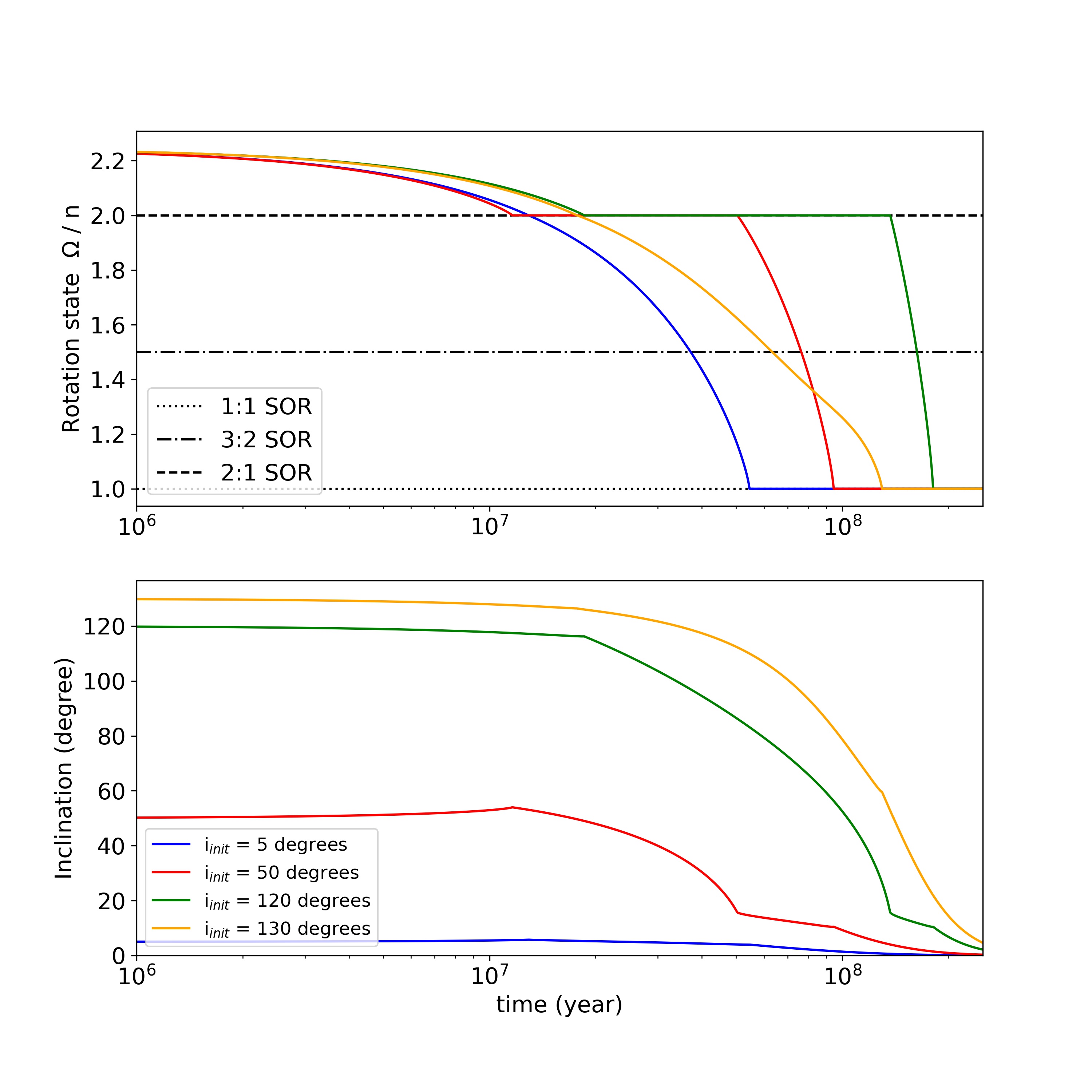}
        \caption{Evolution of a Sun-Venus-like system.
            The top panel shows the evolution of the rotation state in terms of $\Omega /n$ for an initial rotation period of about 100 days.
            The bottom panel shows the evolution of the orbital inclination for different initial inclinations of about $0$, $50$, $120$, and $130$ degrees.
            }
            \label{fig:Espem_spin_incl_evol}
    \end{figure}
    Figure \ref{fig:Espem_spin_incl_evol} shows the evolution of a Venus-like planet with the multilayer internal reference structure (see Sec. \ref{Subsec:Tidal_love_num}) after $2\times 10^8$ years of evolution for three initial inclinations and an initial rotation of $100$ days in a circular orbit with the parameters presented in table~\ref{tab:initial_parameters}.
    For an initial inclination of $5$ degrees (blue curve), the tides act to synchronize the rotation of the planet in $5.5\times 10^7$~years, while the inclination is damped to zero on longer timescales.
    However, the spin can be trapped in SOR if the initial inclination is high enough.
    For the initial inclination of $50$ and $120$ degrees, the planet is captured in the 2:1~SOR for a few $10^7$~yr.
    
\begin{figure*}[h!]%
        \centering
        \includegraphics[width=\textwidth, trim = 0.0cm 0.0cm 0.0cm 0.0cm] {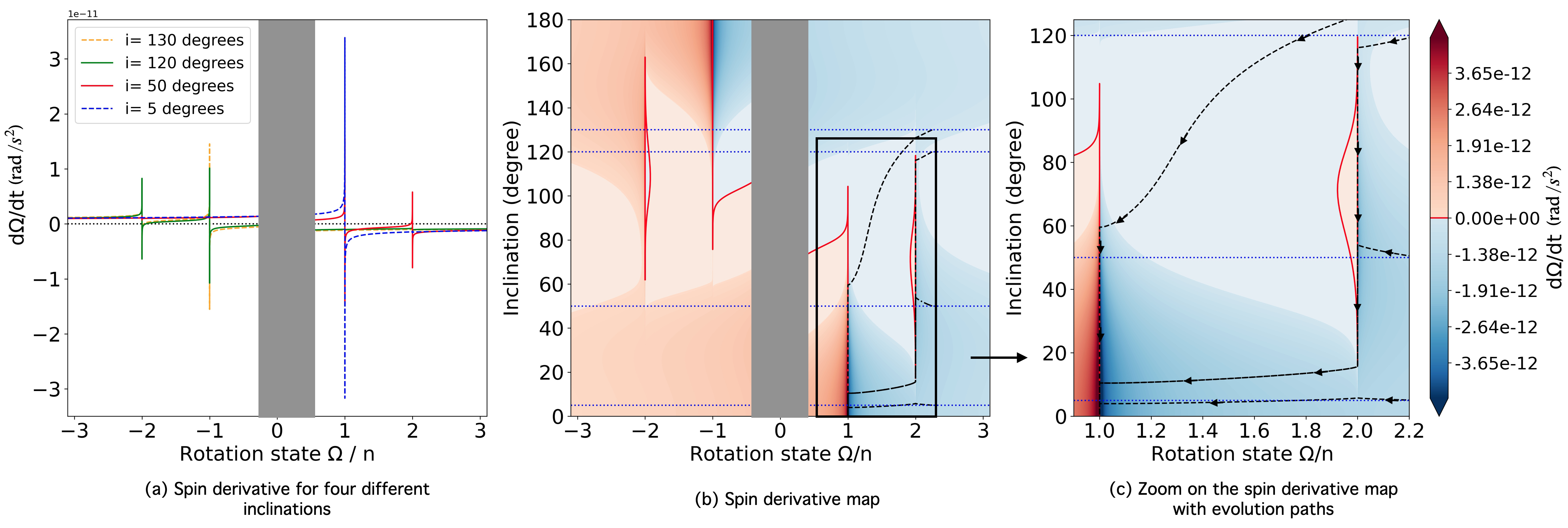}%
        \caption{
            Spin derivative $\mathrm{d}\Omega/\mathrm{d}t$ (in $rad/s^2$).
            The left panel shows the spin derivative as a function of the rotation state $\Omega/n$ and for different inclinations ($5$, $50$, $120$, and $130$~degrees).
            The middle panel represents the value of the spin derivative as a function of $\Omega/n$ and the inclination.
            The red areas depict the positives values, and the blue areas depict the negative values.
            The dotted black lines of the middle and right panels represent the four inclination values ($5$, $50$, $120$, and $130$~degrees) plotted in the left panel.
            The right panel shows a zoom into the square drawn in the middle panel, centered on the $2:1$ SOR.
            The dashed black curves depict the paths of the simulations presented in Fig.~\ref{fig:Espem_spin_incl_evol}.
            The gray areas hide the part of the figure close to the null rotation, where the equations used are no longer valid due to the gyroscopic approximation (see Sec~\ref{Subsec:secular_equations}).}%
        \label{fig:spin_incl_3Panel}
    \end{figure*}

    As in the previous section, we investigated how the derivative of the spin depends on rotation state $\Omega/n$ and the inclination without eccentricity.
    Figure~\ref{fig:spin_incl_3Panel}b shows the spin derivative strength in the plane inclination versus rotation state $\Omega/n$.
    The equilibrium points can be found with the red curves (null values of the derivative).
    In the same manner as in Fig~\ref{fig:spin_ecc_3Panel}, the stable equilibrium states must satisfy the condition of a positive torque (in red) on its left and a negative torque (in blue) on its right for positive values of the rotation state $\Omega/n$, and inversely for negative values of $\Omega/n$.
    Figure~\ref{fig:spin_incl_3Panel}a shows that only one equilibrium is possible for low inclinations: synchronous rotation.
    Increasing the inclination allows other SORs to appear (e.g., the 2:1 SOR). 
    For inclinations higher than $120$ degrees, the prograde rotations are no longer equilibrium points, but the retrograde rotations are, such as the -2:1 SOR at $\Omega/n=-2$.
    A symmetry with respect to a $90$-degree inclination exists.
    This symmetry is clearly visible in the middle panel of Fig~\ref{fig:spin_incl_3Panel}b.
    In particular, the torque at $130$-degree inclination is symmetric of the $50$-degree inclination.
    We highlight that no SORs lie above the 2:1 SOR in rotation.
    Figure~\ref{fig:spin_incl_3Panel}b shows no SORs close to the $\Omega/n = 3 ~\text{or}~ =1.5$.
    Higher spin states were also studied, but as they do not exhibit any SORs. 
    We therefore did not explore a spin rate higher than $100$~days.
    Higher rotation rates require longer timescales to evolve than the present age of the Solar System and are therefore not presented in this paper.

    The evolution paths of Fig.~\ref{fig:Espem_spin_incl_evol} are overplotted in Fig.~\ref{fig:spin_incl_3Panel} for the four initial inclinations of $5$, $50$, $120$, and $130$ degrees.
    For two of these initial inclinations ($50$ and $120$ degrees), we see a capture in the 2:1 SOR (in Fig.~\ref{fig:Espem_spin_incl_evol} and in Fig~\ref{fig:spin_incl_3Panel}b-\ref{fig:spin_incl_3Panel}c).
    This resonance island is stable for inclinations greater than $15$ and lower than $120$ degrees.
    Thus, if the initial spin is higher than $\Omega/n=2$, the spin is always be damped and trapped in the 2:1 SOR (for an inclination between 15 and 120~degrees).
    The spin remains in the 2:1 SOR until the inclination becomes too low to stably maintain this configuration.
    
\begin{figure}[h!]
        \centering
        \includegraphics[width=0.5\textwidth, trim = 0.2cm 0.15cm 0.2cm 0.2cm, clip]{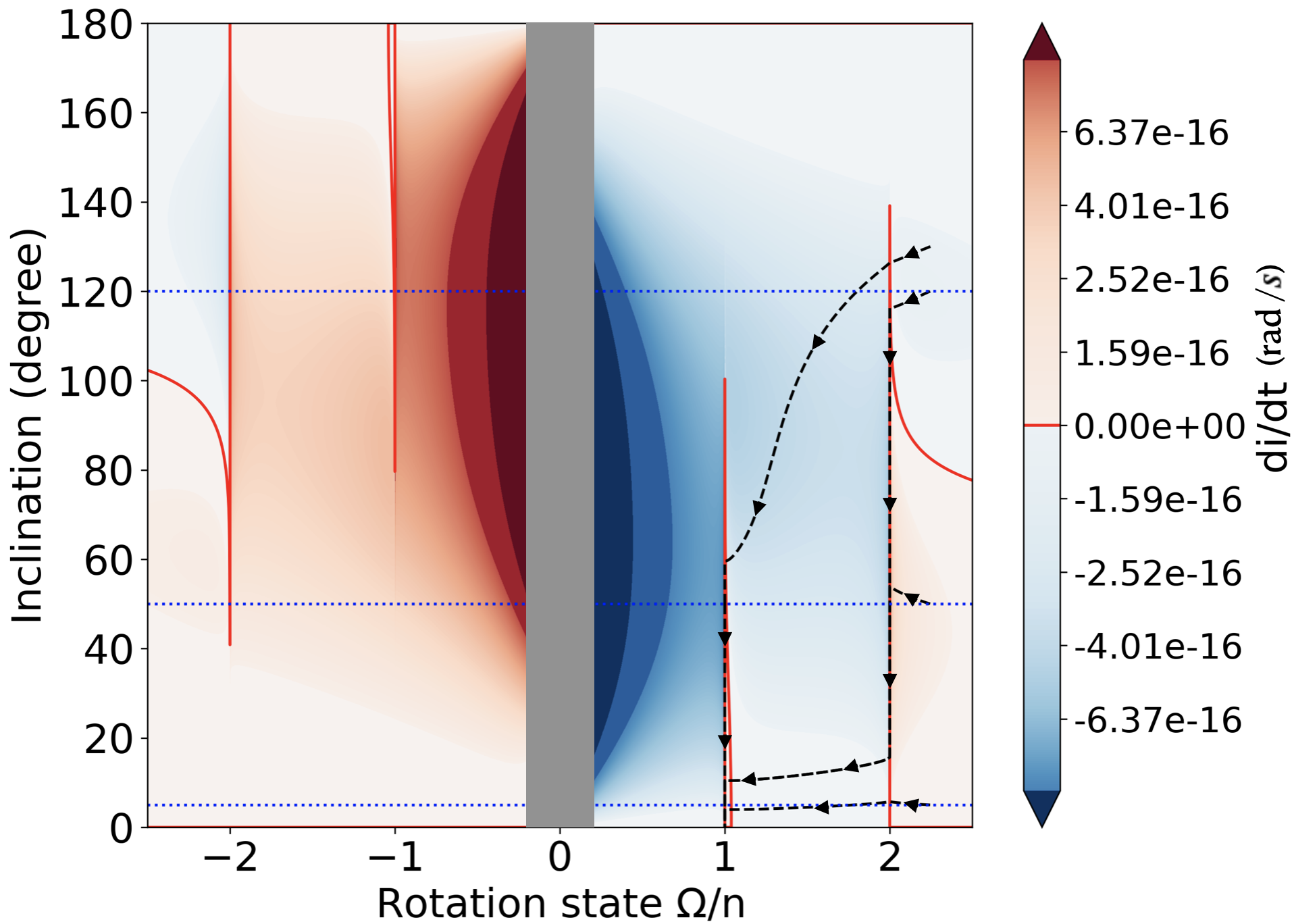}
        \caption{
        Same as Fig~\ref{fig:spin_incl_3Panel}b, with the inclination derivative $\mathrm{d}i/\mathrm{d}t=f(\Omega/n,i)$, in $rad/s$, instead of the spin derivative $\mathrm{d}\Omega/\mathrm{d}t$.}
        \label{fig:2Map_DiDt_NoAtm}
    \end{figure}

    For an initial inclination of $50$~degrees, the inclination appears to be excited by the tides and slightly increases when the spin is higher than the 2:1 SOR.
    This behavior can be explained with the shape of the inclination derivative $\mathrm{d}i/\mathrm{d}t$ plotted in Fig.~\ref{fig:2Map_DiDt_NoAtm}.
    It shows a positive-inclination derivative for a spin higher than the 2:1 SOR and an inclination lower than about $80$~degrees (bottom right corner of the figure).
    It also shows that the inclination should also increase when the spin is slightly higher than the synchronization and for an inclination lower than 100~degrees (red area in the vicinity of the 1:1 SOR).
    This behavior is absent in Fig.~\ref{fig:Espem_spin_incl_evol} because the spin reaches the synchronization very quickly.
    Higher inclination cases can show an interesting behavior.
    Fig.~\ref{fig:spin_incl_3Panel}b clearly shows that for a high initial inclination of about $105$ degrees, the synchronization state (i.e, $\Omega/n =1 $) is no longer a stable configuration.
    Then, if the system starts with a positive rotation and a sufficiently high initial inclination, the spin is damped to the antisynchronization state $\Omega/n =-1$, thus a retrograde rotation.
    As shown in Fig.~\ref{fig:2Map_DiDt_NoAtm}, however, as the spin reaches a negative value, the inclination is driven toward $180$~degrees by the tides.
    This will result in a stable state where the spin is retrograde, in the antisynchronization state, with an inclination of about $180$~degrees.
    This corresponds to a prograde rotation with an orbital inclination of about $0$~degrees. 
    This is consistent with the symmetry on the $\Omega/n=0$ axis and on the $90$-degree axis of the derivative map (Fig.~\ref{fig:spin_incl_3Panel}b).
    
    We investigated spin inclination-driven SORs, such as the 1:1, the 2:1 and their symmetric, the -1:1, and -2:1 SORs, and the evolution of the spin inclination of the planet.
    We show the range of inclination allowing for the SORs from $0$ to $105$~degrees for the $1:1$ and from $20$ to $120$~degrees for the $2:1$~SOR.
    Our simulations in Fig~\ref{fig:Espem_spin_incl_evol} show the particular behavior of the inclination for a rotation rate above $\Omega/n=2$, where the inclination appears to be excited by the tides if the inclination is lower than $80$~degrees.
    Finally, the color maps in Fig.~\ref{fig:spin_incl_3Panel} show the symmetrical properties of the inclination-driven SORs in $\Omega/n=0$ and $i=90$~degrees.
    
    The effect of the thermal tides of a Venus-like atmosphere for different initial spin inclination is studied in the next section.
    
\section{Venus-like atmospheric tides}\label{Sec:Atmopsheric_tides_Results}

    As previous studies showed, the current spin state of Venus cannot be reproduced by involving the solid tides alone \citep{Gold_Soter_1969,DOBROVOLSKIS_1980,Correia_Laskar_2001,Correia_Laskar_2003,correia_long-term_Theory_2003,correia_long-term_NSimulation_2003,Leconte2015}.
    In particular, \cite{Correia_Laskar_2001} showed that atmospheric tides can lead to four final rotation states of Venus, one of which is the retrograde rotation observed today.
    They showed that the current state of Venus cannot be reached for any initial configuration, however.
    We can consider that the current spin inclination of Venus is either high (about $177.36$~degrees) and has a rotation period of $5832.6$ hr, or a low spin inclination (about $2.64$~degrees) and a retrograde rotation.
    We use the case of Venus as a reference.

    The next section (Sec~\ref{subsec:Atmopsheric_tides_Csteparameters}) explores the evolution of a Venus-like planet in the spin and inclination parameter space, with a nonevolving atmosphere, a constant luminosity, and a nonevolving internal profile.
    Section \ref{subsec:Atmopsheric_tides_LuminosityVariation} explores the effect of the luminosity evolution of the host star, accounting for a simple prescription for the atmospheric evolution.
    
    \subsection{Constant luminosity, nonevolving atmosphere}\label{subsec:Atmopsheric_tides_Csteparameters}
    
    \cite{Leconte2015} fit the parameters of their analytical solution of the pressure bulge (Eq.~\ref{eq:pressure_bulge_amplitude}) to their GCM simulation to model the thermal tides.
    These parameters are given in Section~\ref{Subsec:Atmospheric_tides_physics}.

    In this section, we investigate the effect of the thermal forcing produced by the host star on the atmosphere.
    We considered two  models for the interior: a multilayer model (introduced in Section~\ref{Subsec:Tidal_love_num}) and the fitted homogeneous model (introduced in Section~\ref{Subsubsec:Hot_Homogeneous_profile}).
    For the thermal tides, we used the analytical model of thermal tides fit on the present-day Venus (introduced in Section~\ref{Subsec:Atmospheric_tides_physics}.)
    The frequency dependence of the corresponding Love numbers is given in Fig.~\ref{fig:imk2_cold-hot_profile}.
    
    As discussed in Section~\ref{Subsubsec:Hot_Homogeneous_profile}, the solid tides corresponding to the multilayer model and its two variants (the V0.1 and V100 profiles) do not allow a stable-equilibrium point close to the current frequency of Venus as they are either too strong or too weak.
    We fit a homogeneous hot profile, using the method of \cite{Bolmont2020a}, in order to find an equilibrium point close to the Venusian frequency (see  Sec~\ref{Subsubsec:Hot_Homogeneous_profile}).
    The shape of the spin derivative in the bottom panel of Fig.~\ref{fig:Dspin_V1V4Hom_Lecfit} shows one stable equilibrium state and two unstable equilibrium states.
    The two unstable states, close to the synchronization, are also present in the highly viscous V100 profile case.
    The negative stable spin state was fit to correspond to the retrograde state of Venus.
    The $1:1$ synchronous spin state remains stable.
    
\begin{figure*}[h!]%
        \centering
        \includegraphics[width=\textwidth, trim = 0.0cm 0.0cm 0.0cm 0.0cm] {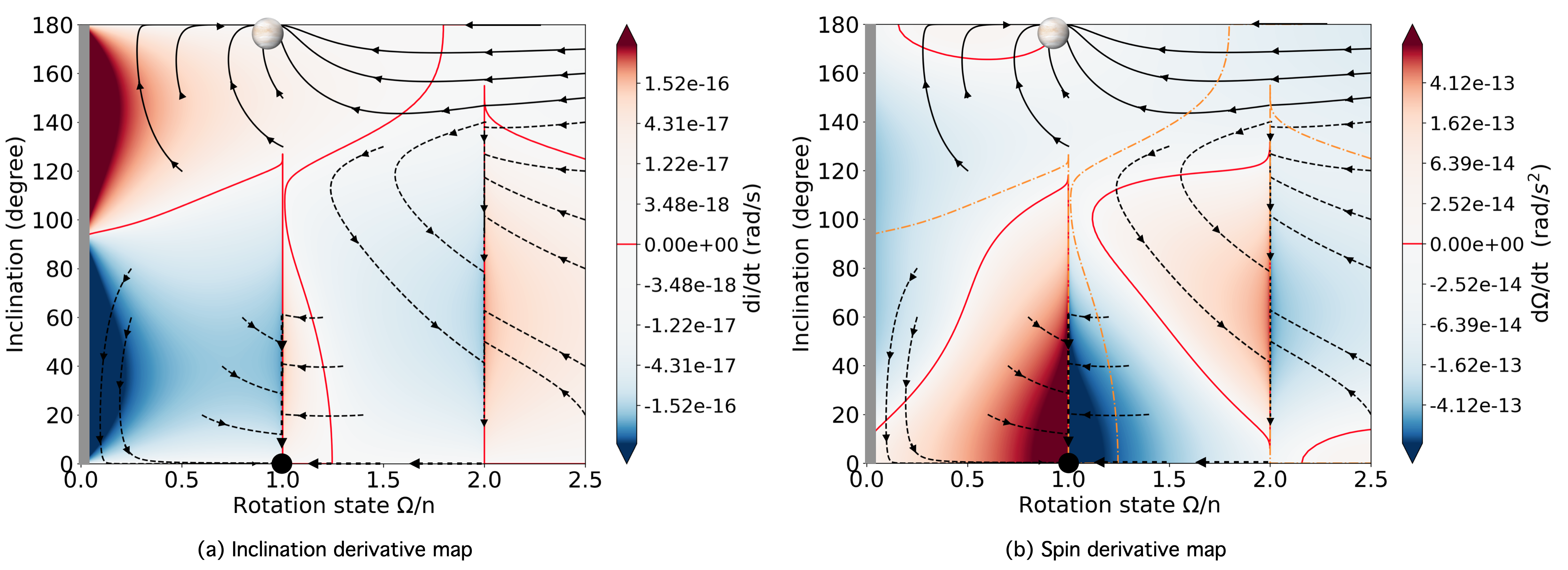}%
        \caption{
            Left panel: Inclination derivative $di/dt$ (in $rad/s$) as a function of the spin $\Omega/n$ ($\Omega$ and $n$ are the planetary spin and mean motion respectively) and for different inclinations (from $0$ to $180$ degrees).
            Right panel: Spin derivative $d\Omega/dt$ (in $rad/s^2$) as a function of the spin $\Omega/n$ and inclinations (from $0$ to $180$ degrees).
            The red lines represent the null derivative in both panels.
            The arrows show the evolution of the system.
            They are consistent with the sign of the inclination derivative (left panel) and the spin derivative (right panel).
            The dotted orange lines represent the null points of the inclination derivative from the left panel (in red in the left panel).
            The image of Venus at the top of the two panels corresponds to the current state of Venus.
            The black dot represents the 1:1 synchronization state.
            The gray area hides the part of the plots close to the null rotation.}
        \label{fig:spin_incl_V1Homogen_Lec_3Panel}
    \end{figure*}

    As in Sec~\ref{Subsec:Solid_tides_InclinedSOR_Low_inclination}, we used the derivative maps of $d\Omega/dt$ and $di/dt$ as a function of $\Omega/n$ and $i$ to represent the evolution of the system in Fig~\ref{fig:spin_incl_V1Homogen_Lec_3Panel}.
    As discussed in Sec~\ref{Subsubsec:Hot_Homogeneous_profile}, we constrained our study at low spin rates.
    Because no inclined SORs are higher than the 2:1 (see Sec~\ref{Subsec:Solid_tides_InclinedSOR_Low_inclination}), the initial spin rate of the simulations was set to $\Omega/n=2.5$ for most of our simulations.
    We can find the set of initial spins and spin inclinations that can lead to a stable state close to the current rotation state of Venus (with an inclination as high as $180$~degrees).
    Figure~\ref{fig:spin_incl_V1Homogen_Lec_3Panel}a and \ref{fig:spin_incl_V1Homogen_Lec_3Panel}b show the evolution of the system through the map, in which each curve represents an ESPEM simulation.
    The solid lines show the cases leading to an inclination as high as $180$~degrees with a prograde rotation, close to the current state of Venus.
    The dashed lines show the cases leading to the $1:1$~SOR and an inclination of $0$~degrees (i.e., a prograde rotation with a null inclination).
    The maps show that the atmospheric tides can drive the system toward the high-inclination state and keep the rotation on the prograde spin rate (i.e., prograde rotation with a high inclination) if the initial spin inclination is higher than about $150$~degrees with a fast initial rotation.
    This configuration can be reached through the effect of a chaotic motion in the Solar System for the case of Venus \citep{Correia_Laskar_2003}.

    \cite{Correia_Laskar_2001} argued that the current state of Venus can be described with four final states, depending on the evolution path of the planet.
    In their work, the paths leading to the current state of Venus either evolved by increasing the spin inclination toward $180$~degrees and keeping the spin on a prograde rotation by either decreasing the spin toward retrograde rotation or keeping the spin inclination to zero degrees.
    In this study, the retrograde rotation can be reached only from the evolution of the spin inclination toward the high-inclination states.
    None of the paths shown in Fig~\ref{fig:spin_incl_V1Homogen_Lec_3Panel}b crosses the null spin state.
    Any positive rotation with a low-inclination configuration will drive the system in the synchronous state.
    As the chaotic effect of a third body will only perturb the spin inclination of the planet it is therefore unlikely that the rotation has crossed the null spin during its evolution given our set of hypotheses.
    
    \subsection{Luminosity variation}\label{subsec:Atmopsheric_tides_LuminosityVariation}
    
\begin{figure}[h!]
        \centering
        \includegraphics[width=0.45\textwidth,trim = 0.0cm 0.0cm 0.5cm 1cm, clip]{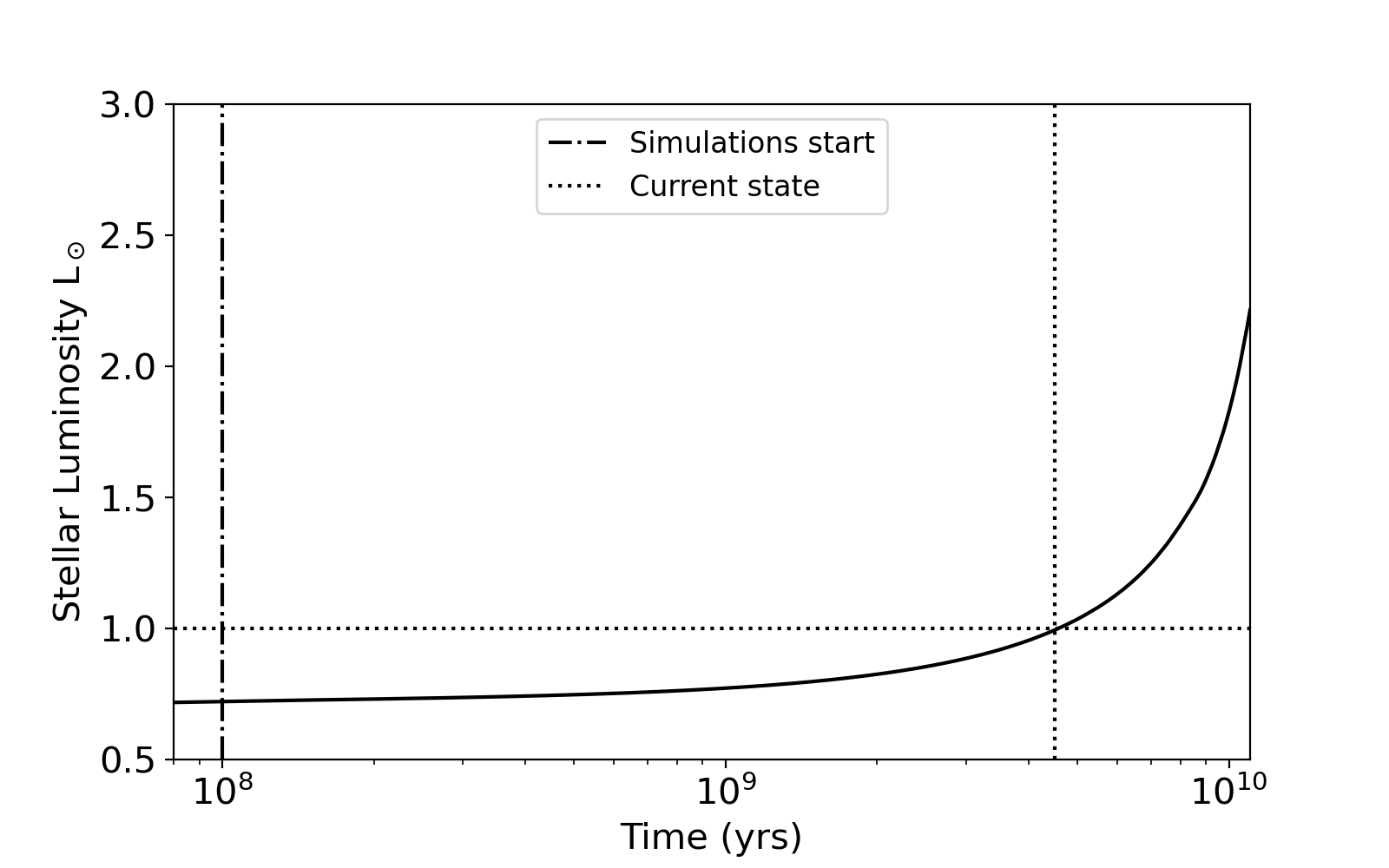}
        \caption{
        Luminosity variation of a Sun-like star over the time from the beginning to the end of the MS.
        The dash-dotted line represents the simulation start time.
        The dotted lines represent the current time state.
        }
        \label{fig:Star_lum_var}
    \end{figure}

    Dynamical studies of the thermal tides \citep{Correia_Laskar_2001,Correia_Laskar_2003,Leconte2015} have considered a constant luminosity. 
    As the thermal forcing depends on the heat flux of the host star, we also investigated the effect of an evolving luminosity on the rotation evolution of a Venus-like planet. 
    The luminosity evolution of the Sun-like star in ESPEM comes from simulations with the stellar evolution code STAREVOL \citep{Amard2016}.
    Figure~\ref{fig:Star_lum_var} shows the luminosity variation of the Sun-like star we considered.
    The thermal Love number was computed with the luminosity dependence with the formulation of \cite{auclair_2017b} as 
    \begin{equation}
        \begin{split}
            \Im(k_2^{atm}(\sigma)) &= - \frac{4}{32}\frac{\kappa \tau \varsigma \epsilon L_\star a}{R_A T_0 M_\star R} \frac{\sigma}{\sigma^2 +\omega^2_0}, \\
            \label{eq:thermal_Lovenumber_auclair}
        \end{split}
    \end{equation}
    with $L_\star$ and $M_\star$ the stellar mass and luminosity respectively, $R$ the radius of the planet, $a$ the semi-major axis, $\tau$ a weight parameter that gives the efficiency of the coupling between the atmosphere and the surface ($0<\tau<1$), $\varsigma$ a shape factor depending on the spatial distribution of tidal heat sources, $\kappa$ the power per mass unit radiated by the atmosphere (where the atmosphere is assumed to behave like a graybody, i.e., Newtonian cooling), $\epsilon$ the effective fraction of power absorbed by the atmosphere, $\sigma$ the excitation frequency, $\omega_0$ the radiative frequency, $T_0$ the equilibrium surface temperature of the atmosphere, $R_A$ the specific gas constant defined as $R_A = R_{GP}/\mathcal{M}_A$ ($R_{GP}$ and $\mathcal{M}_A$ being the perfect gas constant and the mean molar mass respectively), and $\alpha$ the shape factor depending on the spatial distribution of tidal heat sources.
    The values of the parameters we used are presented in table~\ref{tab:thermal_tide_parameters}.
    
    \begin{table}[h!]
        \caption{Numerical values for the thermal Love number of Eq.~\ref{eq:thermal_Lovenumber_auclair}.}
        \label{tab:thermal_tide_parameters}
        \begin{tabular}{lll}
        \hline
        \hline
         Parameter & Values & Units \\
        \hline
         $T_0$ & $737$ & K \\
         $\tau$ & $1$ & - \\ 
         $\varsigma$ & $0.19$ & - \\ 
         $\kappa$ & $0.286$ & - \\ 
         $\epsilon$ & $0.04$ & - \\
         $\omega_0$ & $3.77E-7$ & s$^{-1}$ \\
         $R_{GP}$ & $8.314$ & J mol$^{-1}$ K$^{-1}$ \\
         $\mathcal{M}_A$ & $43.45$ & g mol$^{-1}$ \\
        \hline
        \end{tabular}
    \end{table}
    
    We started the simulation at $100$~Myr regarding the timescale of the rocky planets to form \citep{CHAMBERS_2004}.
    We considered the atmosphere to be fully formed quickly, over the first $1$~Myr after the formation of the planet.
    Then, as the evolution of the atmosphere is very uncertain, we considered it as non-evolving.
    The following part presents the evolution of the system after 3.6~Gyr of evolution.
    
\begin{sidewaysfigure*}[h!]%
        \centering
        \includegraphics[width=\textwidth,] {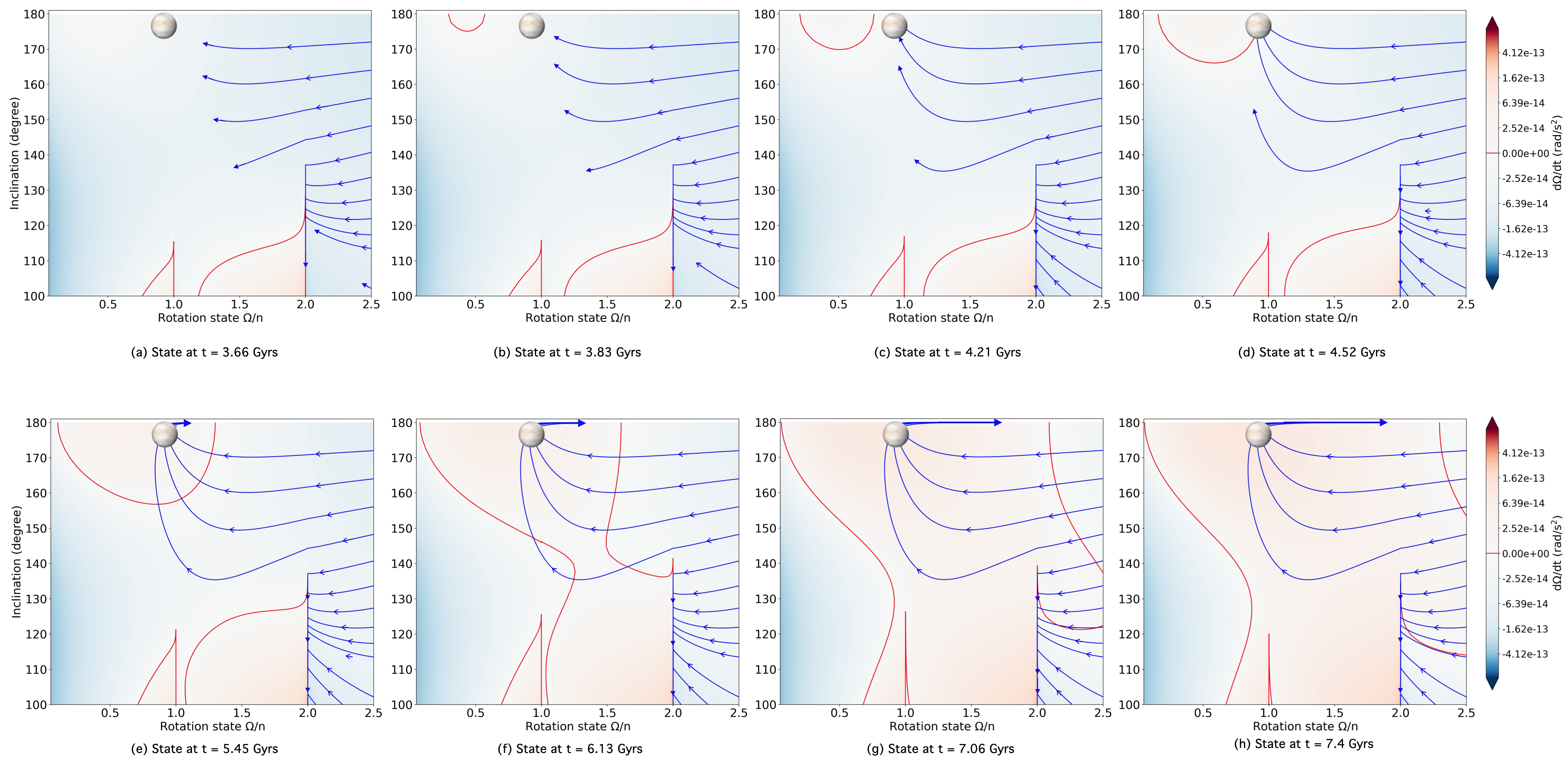}%
        \caption{Same as Fig~\ref{fig:spin_incl_V1Homogen_Lec_3Panel}b for a variable luminosity.
        Each panel shows the derivation map and ESPEM simulation paths at different time steps.
        The image of Venus at the top of the plots correspond to the current spin state of Venus.
        The blue curves correspond to ESPEM simulations evolving with time and luminosity as described in Sec.~\ref{subsec:Atmopsheric_tides_LuminosityVariation}.}
        \label{fig:spin_incl_V1Homogen_LumVar}
    \end{sidewaysfigure*}

    Figure~\ref{fig:spin_incl_V1Homogen_LumVar} shows the spin versus inclination maps with the ESPEM simulation overplotted, in the same manner as the Fig~\ref{fig:spin_incl_V1Homogen_Lec_3Panel}a and \ref{fig:spin_incl_V1Homogen_Lec_3Panel}b.
    Figures~\ref{fig:spin_incl_V1Homogen_LumVar}a to \ref{fig:spin_incl_V1Homogen_LumVar}h show the apparition and the evolution of the equilibrium state close to the current state of Venus (red curve appearing in the top left corner of the map from panel \ref{fig:spin_incl_V1Homogen_LumVar}b to panel \ref{fig:spin_incl_V1Homogen_LumVar}h).
    In the early stages of the simulations, the stellar flux is lower than today, and gravitational tides dominate the thermal ones.
    This means that the spin and inclination evolution is mainly driven by the gravitational tides, following a path consistent with figure~\ref{fig:spin_incl_3Panel}c.
    Figure~\ref{fig:spin_incl_V1Homogen_LumVar}d corresponds to a situation in which an equilibrium close to the current state of Venus is found for the current age of the Solar System.
    We must emphasize that these maps were set to reproduce the steady state at the current state of Venus for the current solar luminosity.
    Figures~\ref{fig:spin_incl_V1Homogen_LumVar}e to \ref{fig:spin_incl_V1Homogen_LumVar}h show that the Solar luminosity increases faster than the spin state.
    Thus, as the luminosity increases, the spin never stays in a stable configuration, but continuously evolves toward the stable state.
    The evolution of the theoretical equilibrium rotation state between the gravitational and thermal tides can be found by finding the rotation rate $\Omega_\text{eq}$, which satisfies $k_2^{\text{grav}}(\sigma_\text{eq})=k_2^{\text{thermal}}(\sigma_\text{eq})$.
    
\begin{figure}[h!]
        \centering
        \includegraphics[width=0.45\textwidth,]{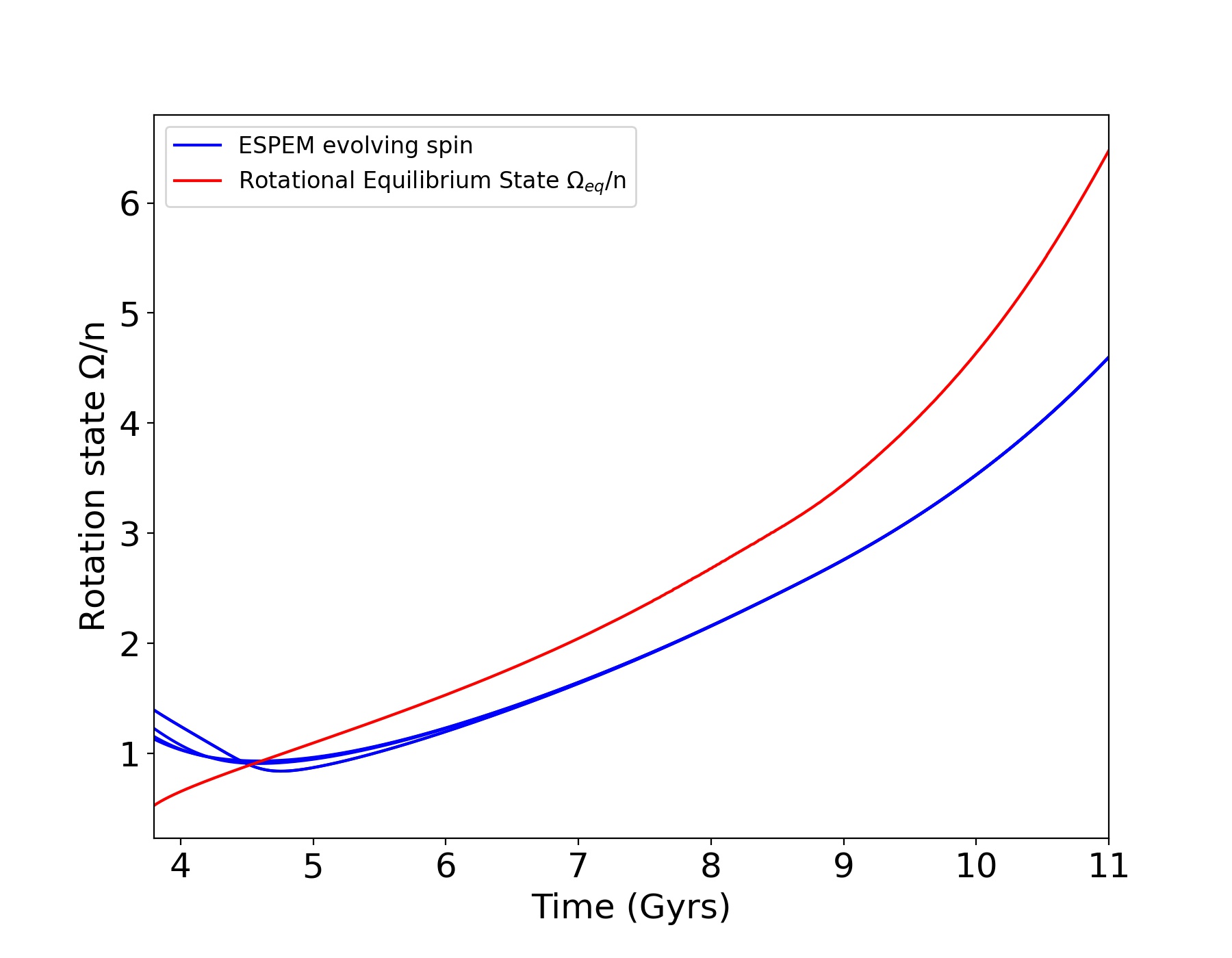}
        \caption{
        Evolution of the rotational equilibrium state $\Omega_{eq}$ (in red) evolving with the stellar luminosity (Fig.~\ref{fig:Star_lum_var}).
        In blue, we show the four curves corresponding to the rotational evolution from ESPEM that crossed the current spin state of Venus in Fig.~\ref{fig:spin_incl_V1Homogen_LumVar}.
        }
        \label{fig:Equilibruim_Rotation_state}
    \end{figure}

    Figure~\ref{fig:Equilibruim_Rotation_state} shows the evolution of the equilibrium rotation rate in terms of $\Omega_\text{eq}/n$ and the evolution of the rotation rate $\Omega/n$ from the ESPEM simulations shown in Fig.~\ref{fig:spin_incl_V1Homogen_LumVar}.
    The equilibrium states were determined numerically by finding the rotation state $\Omega_{eq}$ that verifies the equality between the gravitational and the thermal Love number $\Im(k_2^{\text{grav}}(\Omega_{eq}/n)) = \Im(k_2^{\text{thermal}}(\Omega_{eq}/n))$ over the evolution of the stellar luminosity.
    We show the evolution of the four simulations that crossed the current spin state of Venus during their evolution in Fig.~\ref{fig:spin_incl_V1Homogen_LumVar}.
    These cases crossed the equilibrium state close to the current time, but the equilibrium point evolved faster than the rotational state of the simulation.
    Figure~\ref{fig:spin_incl_V1Homogen_LumVar}h shows that the thermal tides eventually become stronger than the gravitational tides across a large parameter space as the luminosity increases.

    In summary, the luminosity evolution leads to two effects.
    First, the equilibrium changes as the balance between the gravitational tides and thermal tides evolves (thermal tides dominate as the luminosity increases).
    Second, the planet cannot stay in equilibrium because the timescale of the spin evolution is longer than the timescale of the luminosity evolution.
    Concerning the first point, as the luminosity increases and the thermal tides becomes stronger, the equilibrium moves to higher spin rates (farther from synchronization).
    Concerning the second point, the rotation of the planet always chases the equilibria indefinitely (considering a nonevolving atmosphere).
    
\section{Conclusion}\label{Sec:discussion}
 
    We presented the recent implementation of the effect of the tides raised by the star on a telluric Venus-like planet in the code ESPEM.
    We added the secular evolution of the osculating elements of the planetary orbit (a, e, i, $\omega$, and $\Omega$), that is the semi-major axis, the eccentricity, the inclination, the longitude of ascending node, the argument of periastron, and the planetary spin.
    We followed the secular equations published by \cite{Boue_Efroimky_2019}, which describe the evolution of the osculating elements of the orbit of the planet under tidal perturbations following the Kaula formalism \citep{Kaula_1964}.
    Our implementation includes gravitational and thermal tides, which allowed us to study the tidal effect of an arbitrary atmosphere on an arbitrary planet, provided that the tidal Love numbers $k_2$ associated with its atmosphere and internal structure are known.
    
    First we focused in Sec~\ref{Subsec:Solid_tides_EccentricSOR_Low_inclination} on the eccentricity-driven SORs and validate our implementation by finding the 1:1, 3:2, 2:1, and 5:2 SORs, depending on the eccentricity value.
    Our results are consistent with the findings of \cite{Walterova2020}.
    Then we investigated in Sec~\ref{Subsec:Solid_tides_InclinedSOR_Low_inclination} the inclination-driven SORs, as shown by \cite{Boue2016}, here with the Andrade rheology.
    In particular, we find the 1:1, the 2:1 and their symmetric, the -1:1, and -2:1 SORs.

    In Sec~\ref{subsec:Atmopsheric_tides_Csteparameters} we investigated the effect of a thick Venus-like atmosphere with the implementation of the analytical model of \cite{Leconte2015}.
    We used their fit parameters, chosen so that GCM simulations reproduce the current state of Venus.
    We fit a homogeneous internal structure that allowed gravitational tides to balance thermal tides at the frequency of Venus.
    We must emphasize that the de-spinning of Venus is a difficult task.
    Then, we constrained our work to lower initial rotation rates.
    We find that depending on the initial spin rate and initial spin inclination, either a spin inclination of about zero in the synchronization state or a state close to the current retrograde rotation of Venus results, with a spin rate close to the synchronization and a spin inclination of $180$~degrees.
    The synchronization state (1:1~SOR) is reached when the planet starts with a spin inclination lower than about $120$ degrees and a prograde spin in our simulations.
    The latter state can be reached when the planet starts either with a high spin inclination (higher than about $120$ degrees) and a prograde rotation, or with a spin inclination lower than about $60$ degrees and a retrograde rotation.
    Our results are consistent with the final spin state of Venus found by \cite{Correia_Laskar_2001}, who computed the evolution of the spin and obliquity of Venus under the solar tides.
    We point out, however, that \cite{Correia_Laskar_2001} used different models for the gravitational tides, which less appropriate than an Andrade rheology for silicate bodies \citep[i.e., the CTL,][]{Hut1981, Goldreich1966,Efroimsky2013} and they included the core-mantle friction.
    The core-mantle friction helps to damp the spin inclination of the planet \citep{Correia_Laskar_2001} and should help the gravitational tides to balance the thermal tides.
    Furthermore, they also accounted for the chaotic motion in the Solar System \citep{Laskar1990}.
    In particular, they reported that the chaotic motion helps to transition from low to high inclination.
    We cannot reproduce this with only two bodies.
    Further developments of the ESPEM code will include these effects.
    
    In Sec~\ref{subsec:Atmopsheric_tides_Csteparameters} we assumed that the spin state of Venus was in equilibrium state to fit an internal model to the thermal tides.
    Our results in Sec~\ref{subsec:Atmopsheric_tides_LuminosityVariation} showed, however, that this may not be the case, and the spin of Venus may still be evolving because of the variation in the solar luminosity.
    Thus, we investigated the effect of the evolving luminosity on the thermal tides.
    The evolution of the stellar luminosity leads to a continuous change in the balance between the gravitational and the thermal tides.
    The rotation of the planet will then continually increase as the luminosity increases.
    The luminosity evolving faster than the spin, the rotation of the planet will chase the equilibrium state without reaching it.
    We must highlight that the way in which the thermal tides will continue to increase as the rotation of the planet increase is unclear.
    In our case, the Maxwell-like frequency-dependent model of thermal tides overestimates the strength of the thermal tides at high frequencies, as the model was fit for the low-frequency regime \citep{Auclair_2019}.
    Further studies with GCM simulations of Venus at higher rotation rates will help to answer this question.
    

    Exploring the complete evolution of the rotation of Venus requires calculating the dynamical evolution of the planet in (at least) three-body simulations.
    The effect of the evolution of the internal structure and the atmosphere of the planet must also be investigated.
    The strength of the gravitational and thermal tides, and thus the balance between the two contributions, should have varied strongly during the evolution of the planet since its formation.
    The current internal structure and atmospheric tides are not sufficiently constrained, however.
    Future missions to Venus, such as EnVision \citep{EnVission_mission_2020}, DAVINCI \citep{Garvin_2022_DAVINCI}, and VERITAS \citep{VERITAS_mission}, will bring valuable data on the internal state of Venus and on the thermal atmospheric response of the planet \citep{Bills_2020}.
    They will help to determine whether the planet is in equilibrium between the gravitational and the thermal tide. 
    These constraints could also help reconstruct the thermal evolution of the planet, which would impact the competition between the gravitational and thermal tides and thus the rotational evolution. 
    Finally, better observations of the atmosphere, together with additional modeling of the Venusian atmosphere, would help constrain the thermal tide. 
    In particular, estimating the response of the atmosphere with a GCM to other frequencies would be extremely helpful.
    
    In the context of exoplanets, we need to consider a relevant model of tides for rocky exoplanets to characterize their surface and potential habitability.
    We have shown that a relevant tidal model for rocky planet allows a higher spin state than the synchronization, such as eccentricity-driven SORs and also inclination-driven SORs.
    Planets on a large orbit can keep nonzero eccentricity or obliquity because they evolve on a longer timescale, and can still be trapped in this eccentricity or obliquity-driven SORs.
    When a planet has an atmosphere, thermal tides can excite the spin inclination to high values because thermal tides drive the spin of Venus in its current state through the chaotic motion of the Solar System.
    The strength of the thermal tides also depends on the surface pressure, and thus on the total mass of the atmosphere, on the composition that determines the atmospheric absorption, and on the dynamics of the atmosphere.
    These dependences should be investigated in future studies.
    We showed that the variation in host star luminosity can also prevent the rotation of a planet from reaching equilibrium  between gravitational and thermal tides.
    This behavior must be further studied for different types of star, that is, different radiation spectra, and with more elaborate models of thermal tides that take the wavelength dependence of the irradiation and the composition of the atmosphere into account.
    The new generation of instruments, that is, the JWST and ARIEL \citep{Greene2016_JWST,Tinetti_2021_ARIEL, Edwards_2022_ARIEL_target}, will provide valuable data on the atmosphere of rocky worlds.
    The correct modeling of the dynamical state of exoplanets is then crucial to constrain their surface condition.

\begin{acknowledgements}
    This work has been carried out within the framework of the NCCR PlanetS supported by the Swiss National Science Foundation under grants 51NF40\_182901 and 51NF40\_205606. The authors acknowledge the financial support of the SNSF (grant number: 200021\_197176).
    All the members from CEA acknowledge support from GOLF and PLATO CNES grants of the Astrophysics Division at CEA.
    The computations were performed at University of Geneva on the Baobab and Yggdrasil clusters.
    This research has made use of NASA's Astrophysics Data System.
    The authors thank Drs. Jérémy Leconte, Pierre Auclair-Desrotour and Gwenaël Boué for interesting discussions about the thermal tides.
\end{acknowledgements}

%
%

\bibliographystyle{aa}
\bibliography{Biblio}

\begin{appendix} 

\section{Secular equations}\label{appendix:secular-equation}

    This section presents the secular equations we implemented in the code ESPEM.
    The secular equations used were developed by \cite{Boue_Efroimky_2019}, who revisited the secular equations of \cite{Kaula_1964}.
    We used their equations $116$ to $123$, which describe the secular equations of the osculating elements, $a$, $e$, $i$, $\bar{\Omega}$, $\omega$, and $\varepsilon$ the semi-major axis, the eccentricity, the inclination, the longitude of ascending node, the argument of pericenter and the inclination of the spin axis respectively, and $\dot\theta$ the spin rate.
    Here, the inclination $i$ is defined as the angle between the orbital plane and the planet equator.
    The inclination of the spin axis $\varepsilon$ is defined as the inclination of the spin with respect to the inertial frame.
    
    These equations were computed within the gyroscopic approximation, which implies that the spin rate of a body is much faster than the evolution of the spin-axis orientation.
    This approximation means that considering the limit within which the spin tends to zero cannot be included. 

    Hereafter, the star is taken as the secondary (subscript $_\star$), and the planet as the primary (subscript $_p$), $C$ is the inertia momentum, that is, $C=M_p R_p^2 r_g$, with $r_g$ the gyration radius (which represents the internal density distribution).
    $\beta$ is the reduced mass $\beta = M_p M_\star /(M_p+M_\star)$ and $\mathcal{G}$ the universal gravitational constant.
    $F_{lmp}(i)$ and $G_{lpq}(e)$ are the inclination and eccentricity polynomials respectively (see Appendix~\ref{appendix:table_Kaula}).
    The tidal frequency is defined as $\sigma_{lmpq} = (l-2p+q)n -m\dot\theta$ (with $\dot{\theta}$ and $n$ the spin rate and the mean motion respectively).
    The phases $\nu_{lmpq}$ are defined as $\nu_{lmpq} = (l-2p)\omega + (l-2p+q)\mathcal{M} +m\bar{\Omega}$, with $\mathcal{M}$ the mean anomaly.
    $k_l$ is the modulus of the complex Love number $|\bar{k_l}|$ of degree $l$.
    The tidal potential energy $V_1$ is defined from the Hamiltonian formalism as \citep{Boue_Efroimky_2019} $\mathcal{H} = \mathcal{H_0} +V_1$.
    The tidal perturbing potential $\mathcal{R}$ is computed with $\mathcal{R} = -V_1/\beta$.
    The perturbing potential $U$ (Eq.~\ref{equ:Potential_fourier_modes}) within the Darwin-Kaula formalism can be related to $\mathcal{R}$ with $\mathcal{R} = -\frac{M_\star}{\beta} U$ \citep{Boue_Efroimky_2019}.
    Then, the tidal perturbing potential $\mathcal{R}$ is expressed in the formalism of \cite{Kaula_1964} as
    \begin{equation}
        \begin{split}
            \mathcal{R}(\mathbf{r} , \mathbf{r'}) &= \sum_{l=0}^{+\infty}\Big(\frac{R_p}{a'}\Big)^{l+1} \frac{\mathcal{G}M'}{a'}\Big(\frac{R_p}{a'}\Big)^{l} \sum_{m=0}^{l}\frac{(l-m)!}{(l+m)!}(2-\delta_{0,m}) \\
            & \sum_{p=0}^{l}F_{lmp}(i')\sum_{q=-\infty}^{\infty}G_{lpq}(e') \sum_{h=0}^{l}F_{lmh}(i)\sum_{j=-\infty}^{\infty}G_{lhj}(e) \\
            & k_l(\sigma_{lmpq})\cos{\big[(\nu'_{lmpq}-m\dot\theta')-(\nu_{lmhj}-m\dot\theta)-\varepsilon_l(\sigma_{lmpq})\big]},
        \end{split}
    \end{equation}
    where the subscript $'$ (i.e., $\mathbf{r'}$, $a'$, $e'$, $i'$, $\bar{\Omega}'$, and $\omega'$) corresponds to the coordinate of the perturber, and the parameters without subscript (i.e. $\mathbf{r}$, $a$, $e$, $i$, $\bar{\Omega}$, and $\omega$) correspond to the coordinate where the tides are evaluated.
    In the following, both the perturber and the body on which the tides retro-act are considered to be the star, and therefore we neglect the subscripts.
    All the following equations come from the Hamiltonian development and the method of \cite{Boue_Efroimky_2019}.
    
    The spin derivative equation is expressed from the Hamiltonian formalism as \citep{Boue_Efroimky_2019}
    \begin{equation}\label{appendix:equation-spin_a}
        \begin{split}
            \frac{d^2\theta}{dt^2}\Bigg|_{l=2} = -\frac{\beta}{C}\frac{\partial\mathcal{R}}{\partial\bar{\Omega}}& \\
        \end{split}
    \end{equation}
    We carry out the derivative of the perturbing function $\mathcal{R}$, and the secular equation of the spin derivative is written after some algebra as
    \begin{equation}\label{appendix:equation-spin_b}
        \begin{split}
            \frac{d^2\theta}{dt^2}\Bigg|_{l=2} = -\frac{\mathcal{G}M_\star^2 R^5}{a^6 C} &\sum_{m=0}^{2}m\frac{(2-m)!}{(2+m)!}(2-\delta_{0,m}) \sum_{p=0}^{2}{F_{2mp}(i)}^2 \\ 
            &\sum_{q=-7}^{7}{G_{2pq}(e)}^2 \Im\big(k_2(\sigma_{2mpq})\big) \\
        \end{split}
    \end{equation}
    
    The eccentricity derivative equation is computed from the Hamiltonian equation as
    \begin{equation}\label{appendix:equation-ecc_a}
        \begin{split}
            \frac{de}{dt}\Bigg|_{l=2} &= \frac{1-e^2}{na^2e}\frac{\partial\mathcal{R}}{\partial\mathcal{M}} -\frac{\sqrt{1-e^2}}{na^2e}\frac{\partial\mathcal{R}}{\partial\omega} \\
        \end{split}
    \end{equation}
    Then, the equation of the eccentricity derivative is computed with the eccentricity squared $e^2$ to avoid singularities when the eccentricity tends to zero.
    Thus, we find
    \begin{equation}\label{appendix:equation-ecc_b}
        \begin{split}
            \frac{de^2}{dt}\Bigg|_{l=2} &= -2\sqrt{(1-e^2)}\sqrt{G(M_p+M_\star)}\frac{M_\star}{M_p}\Big(\frac{R_p^5}{a^{13/2}}\Big) \\
                & \sum_{m=0}^{2}\frac{(2-m)!}{(2+m)!}(2-\delta_{0,m}) \sum_{p=0}^{2}F_{2mp}(i)^2 \\ 
                &\sum_{q=-7}^{7}G_{2pq}(e)^2 \Big(\sqrt{1-e^2}(2-2p+q)-(2-2p)\Big) \Im\big(k_2(\sigma_{2mpq})\big) \\ 
        \end{split}
    \end{equation}
    
    The equation of the inclination derivative is also defined from the Hamiltonian formalism as 
    \begin{equation}\label{appendix:equation-incl_a}
        \begin{split}
            \frac{di}{dt}\Bigg|_{l=2} &= \frac{\beta}{C \dot{\theta}\sin{i}} \Big( \frac{\partial\mathcal{R}}{\partial\omega} -\cos{i}\frac{\partial\mathcal{R}}{\partial\bar{\Omega}}\Big) \\
            &-\frac{1}{na^2\sqrt{1-e^2}\sin{i}}\Big( \frac{\partial\mathcal{R}}{\partial\bar{\Omega}} -\cos{i}\frac{\partial\mathcal{R}}{\partial\omega}\Big)\\
        \end{split}
    \end{equation}
    Then, we carry on the derivative of the perturbing function $\mathcal{R}$, and the equation can be written as
    \begin{equation}\label{appendix:equation-incl_b}
        \begin{split}
            \frac{di}{dt}\Bigg|_{l=2} &= \frac{1}{\sin{i}}\frac{M_\star}{M_p} \Big(\frac{R_p}{a}\Big)^{5} \sum_{m=0}^{2}\frac{(2-m)!}{(2+m)!}(2-\delta_{0m}) \times\\
            &\sum_{p=0}^{2} \Big[\frac{\beta n^2 a^2}{C\dot\theta}\Big(m\cos{i} -(2-2p)\Big) -\frac{n}{\sqrt{1-e^2}}\Big((2-2p)\cos{i} -m\Big) \Big] \times\\
            &{F_{2mp}(i)}^2 \sum_{q=-7}^{7} G_{2pq}(e)^2 \Im\big(k_2(\sigma_{2mpq})\big) \\
        \end{split}
    \end{equation}
    
    The equation of the longitude of the ascending node derivative is given as 
    \begin{equation}\label{appendix:equation-long_asc_a}
        \begin{split}
            \frac{d\bar{\Omega}}{dt}\Bigg|_{l=2} &= \Big(\frac{\beta\cos{i}}{C\dot\theta\sin{i}} -\frac{\beta\cos{\varepsilon}\cos{\bar{\Omega}}}{C\dot\theta\sin{\varepsilon}}+\frac{1}{na^2\sqrt{1-e^2}\sin{i}} \Big) \frac{\partial\mathcal{R}}{\partial i} \\
            &+\frac{\beta\cos{\varepsilon}\sin{\bar{\Omega}}\cot{i}}{C\dot\theta\sin{\varepsilon}}\frac{\partial\mathcal{R}}{\partial\bar{\Omega}} -\frac{\beta\cos{\varepsilon}}{C\dot\theta\sin{\varepsilon}}\frac{\sin{\bar{\Omega}}}{\sin{i}}\frac{\partial\mathcal{R}}{\partial\omega} \\
        \end{split}
    \end{equation}
    Then, after some algebra, the equation can be written as 
    \begin{equation}\label{appendix:equation-long_asc_b}
        \begin{split}
            &\frac{d\bar{\Omega}}{dt}\Bigg|_{l=2} =\frac{\mathcal{G}M_\star^2 R_p^5}{a^6 } \sum_{m=0}^{2}\frac{(2-m)!}{(2+m)!}(2-\delta_{0,m}) \Bigg\{ \\
            &\Big(\frac{1}{C\dot\theta\tan{i}} -\frac{\cos{\bar{\Omega}}}{C\dot\theta\tan{\varepsilon}}+\frac{1}{\beta na^2\sqrt{1-e^2}\sin{i}} \Big) \times \\
            &\frac{1}{2} \sum_{p=0}^{2}\frac{\partial~F_{2mp}(i)^2}{\partial~i}\sum_{q=-7}^{7}G_{2pq}(e)^2  \Re\big(k_2(\sigma_{2mpq})\big) \\
            &-\frac{\sin{\bar{\Omega}}\cot{i}}{C\dot\theta\tan{\varepsilon}} m \sum_{p=0}^{2}F_{2mp}(i)^2\sum_{q=-7}^{7}G_{2pq}(e)^2 \Im\big(k_2(\sigma_{2mpq})\big) \\
            &+\frac{1}{C\dot\theta\tan{\varepsilon}}\frac{\sin{\bar{\Omega}}}{\sin{i}} \sum_{p=0}^{l}(2-2p)F_{2mp}(i)^2\sum_{q=-7}^{7}G_{2pq}(e)^2 \Im\big(k_2(\sigma_{2mpq})\big) \Bigg\} \\
        \end{split}
    \end{equation}
    
    The equation of the argument of the periastron derivative is defined as 
    \begin{equation}\label{appendix:equation-arg_peri_a}
        \begin{split}
            \frac{d \omega}{dt}\Bigg|_{l=2} &= -\frac{\beta}{C\dot\theta\sin{i}}\frac{\partial\mathcal{R}}{\partial i} +\frac{\sqrt{1-e^2}}{na^2e}\frac{\partial\mathcal{R}}{\partial e} -\frac{\cos{i}}{na^2\sqrt{1-e^2}\sin{i}}\frac{\partial\mathcal{R}}{\partial i} \\
        \end{split}
    \end{equation}
    Then, the equation can be written as 
    \begin{equation}\label{appendix:equation-arg_peri_b}
        \begin{split}
            &\frac{d \omega}{dt}\Bigg|_{l=2} = \frac{\mathcal{G}M_\star^2 R_p^5}{a^6} \sum_{m=0}^{2}\frac{(2-m)!}{(2+m)!}\frac{(2-\delta_{0,m})}{2} \Bigg[ \\
            &-\Bigg(\frac{1}{C\dot\theta\sin{i}}+\frac{1}{na^2\sqrt{1-e^2}\tan{i}}\frac{1}{\beta}\Bigg) \sum_{p=0}^{2}\frac{\partial~F_{2mp}(i)^2}{\partial~i} \sum_{q=-7}^{7}G_{2pq}(e) \Re\big(k_2(\sigma_{2mpq})\big)\\
            &+\frac{\sqrt{1-e^2}}{na^2e}\frac{1}{\beta} \sum_{p=0}^{2}F_{2mp}(i)\sum_{q=-7}^{7}\frac{\partial~G_{2pq}(e)^2}{\partial~e} \Re\big(k_2(\sigma_{2mpq})\big) \Bigg] \\
        \end{split}
    \end{equation}
    
    Finally, the equation of the inclination of the spin axis is defined as 
    \begin{equation}\label{appendix:equation-obliquity_a}
        \begin{split}
            \frac{d \varepsilon}{dt} &= -\frac{\beta}{C \dot\theta}\Big( \cos{\bar{\Omega}}\cot{i}\frac{\partial \mathcal{R}}{\partial\bar{\Omega}} +\sin{\bar{\Omega}}\frac{\partial\mathcal{R}}{\partial i} -\frac{\cos{\bar{\Omega}}}{\sin{i}}\frac{\partial\mathcal{R}}{\partial\omega} \Big) \\
        \end{split}
    \end{equation}
    Then, after some algebra, we find 
    \begin{equation}\label{appendix:equation-obliquity_b}
        \begin{split}
            \frac{d \varepsilon}{dt} &= \frac{\mathcal{G}M_\star^2 R^5}{a^6 C \dot\theta} \sum_{m=0}^{2}\frac{(2-m)!}{(2+m)!}(2-\delta_{0,m}) \Bigg( \\
                & m \cos{\bar{\Omega}}\cot{i} \sum_{p=0}^{2}F_{2mp}(i)^2\sum_{q=-2}^{2}G_{2pq}(e)^2 \Im\big(k_2(\sigma_{2mpq})\big)\\
                & -\frac{1}{2} \sin{\bar{\Omega}} \sum_{p=0}^{2}\frac{\partial F_{2mp}(i)^2}{\partial i}\sum_{q=-\infty}^{\infty} G_{2pq}(e)^2  \Re\big(k_2(\sigma_{2mpq})\big) \\
                & -\frac{\cos{\bar{\Omega}}}{\sin{i}} \sum_{p=0}^{2}(2-2p)F_{2mp}(i)^2\sum_{q=-2}^{2}G_{2pq}(e)^2 \Im\big(k_2(\sigma_{2mpq})\big) \Bigg) \\
        \end{split}
    \end{equation}
    
\section{Table of the inclination eccentricity from the Fourier development by Kaula, Cayley etc}\label{appendix:table_Kaula}
    The inclination and eccentricity polynomials, $F_{lmp}(i)$ and $G_{lpq}(e)$, respectively, are given by Eqs.~$20$, $23$, and $24$ of \cite{Kaula_1961} and are presented in Tables~\ref{tab:F_inclinaison} and \ref{tab:G_eccentricite}.
    The eccentricity functions are elliptic expansions that can be computed with the Hansen function $\mathbf{X}_{l-2p+q}^{-(l-1),(l-2p)}$ \citep{Tisserand1889}.
    These expansions are discussed by \cite{Izsak1964}.
    
    We considered eccentricities up to $0.3$, which allowed us to consider the eccentricity expansions up to order $7$ \citep[see tables][]{Cayley_Tables-of-dvlp_1860}.
    As the tidal interactions are computed at the quadupolar order $l=2$, the index $m$ is constrained between $0$ and $2$ and $p$ between $-7$ and $7$.
    
    \begin{table}[h]
        \caption{Inclination polynomials $F_{lmp}(i)$ for $l=2$ (\cite{Kaula_1964}) }
        \label{tab:F_inclinaison}
         \centering
         \begin{tabular}{| c | c | c || l | }
            \hline
                l & m & p & $F_{lmp}(i)$ \\
       \hline
       2 & 0 & 0 & $-\frac{3}{8}\sin{i}^2$               \\
       2 & 0 & 1 & $\frac{3}{4}\sin{i}^2 -\frac{1}{2}$   \\
       2 & 0 & 2 & $-\frac{3}{8}\sin{i}^2$               \\
       \hline
       2 & 1 & 0 & $\frac{3}{4}\sin{i} (1+\cos{i}) $ \\
       2 & 1 & 1 & $-\frac{3}{2}\sin{i}\cos{i}$      \\
       2 & 1 & 2 & $\frac{3}{4}\sin{i}(\cos{i} -1)$  \\
       \hline
       2 & 2 & 0 & $\frac{3}{4}(1 +\cos{i})^2$      \\
       2 & 2 & 1 & $\frac{3}{2}\sin{i}^2$           \\
       2 & 2 & 2 & $\frac{3}{4}(1 -\cos{i})^2$      \\
       \hline  
        \end{tabular}
    \end{table}
    
    \begin{table}[h]
        \caption{Eccentricity polynomials $G_{lpq}(e)$ from \citet[]{Cayley_Tables-of-dvlp_1860}, up to order $7$ in eccentricity.}
        \label{tab:G_eccentricite}
         \centering
         \begin{tabular}{| c | c | c || l |}
            \hline
               l & p & q & $G_{lpq}(e)$ \\
       \hline
       2 & 0 & -7 & $ \frac{15625.0}{129024.0}e^7 $  \\
       2 & 0 & -6 & $ \frac{4.0}{45.0}e^6 $ \\
       2 & 0 & -5 & $ \frac{81.0}{1280.0}e^5 +\frac{81.0}{2048.0}e^7  $ \\
       2 & 0 & -4 & $ \frac{1.0}{24.0}e^4 +\frac{7.0}{240.0}e^6 $ \\
       2 & 0 & -3 & $ \frac{1.0}{48.0}e^3 +\frac{11.0}{768.0}e^5 +\frac{313.0}{30720.0}e^7 $    \\
       2 & 0 & -2 & $0 $  \\
       2 & 0 & -1 & $-\frac{1}{2}e +\frac{1}{16}e^3 -\frac{5}{384}e^5 -\frac{143}{18432}e^7$     \\
       2 & 0 & 0 & $ 1 -\frac{5}{2}e^2 +\frac{13}{16}e^4 -\frac{35}{288}e^6 $   \\
       2 & 0 & 1 & $ \frac{7}{2}e -\frac{123}{16}e^3 +\frac{489}{128}e^5 -\frac{1763}{2048}e^7 $   \\
       2 & 0 & 2 & $ \frac{17}{2}e^2 -\frac{115}{16}e^4 +\frac{601}{48}e^6$   \\
       2 & 0 & 3 & $ \frac{845.0}{48.0} e^3 -\frac{32525.0}{768.0} e^5 +\frac{208225.0}{6144.0} e^7 $    \\
       2 & 0 & 4 & $ \frac{533.0}{16.0} e^4 -\frac{13827.0}{160.0} e^6 $         \\
       2 & 0 & 5 & $ \frac{228347.0}{3840.0} e^5 -\frac{3071075.0}{18432.0} e^7 $   \\
       2 & 0 & 6 & $ \frac{73369.0}{720.0} e^6 $         \\
       2 & 0 & 7 & $ \frac{12144273.0}{71680.0} e^7 $    \\
       \hline
       2 & 1 & -7 & $G_{217}(e)$   \\
       2 & 1 & -6 & $G_{216}(e)$   \\
       2 & 1 & -5 & $G_{215}(e)$   \\
       2 & 1 & -4 & $G_{214}(e)$   \\
       2 & 1 & -3 & $G_{213}(e)$   \\
       2 & 1 & -2 & $G_{212}(e)$   \\
       2 & 1 & -1 & $G_{211}(e)$    \\
       2 & 1 & 0 & $ (1 -e^2)^{-3/2} \simeq 1 +\frac{3}{2}e^2 +\frac{15}{8}e^4 +\frac{35}{16}e^6 +\mathcal{O}(e^9)$    \\
       2 & 1 & 1 & $ \frac{3.0}{2.0}e +\frac{27.0}{16.0}e^3 +\frac{261.0}{128.0}e^5 +\frac{14309.0}{6144.0}e^7$    \\
       2 & 1 & 2 & $ \frac{9.0}{4.0}e^2 +\frac{7.0}{4.0}e^4 +\frac{141.0}{64.0}e^6$    \\
       2 & 1 & 3 & $\frac{53.0}{16.0}e^3 +\frac{393.0}{256.0}e^5 +\frac{24753.0}{10240.0}e^7 $  \\
       2 & 1 & 4 & $\frac{77.0}{16.0}e^4 +\frac{129.0}{160.0}e^6 $  \\
       2 & 1 & 5 & $\frac{1773.0}{256.0}e^5 -\frac{4987.0}{6144.0}e^7 $  \\
       2 & 1 & 6 & $\frac{3167.0}{320.0}e^6 $   \\
       2 & 1 & 7 & $\frac{432091.0}{30720.0}e^7 $   \\
       \hline
       2 & 2 & -7 & $G_{207}(e)$  \\
       2 & 2 & -6 & $G_{206}(e)$  \\
       2 & 2 & -5 & $G_{205}(e)$  \\
       2 & 2 & -4 & $G_{204}(e)$  \\
       2 & 2 & -3 & $G_{203}(e)$  \\
       2 & 2 & -2 & $G_{202}(e)$  \\
       2 & 2 & -1 & $G_{201}(e)$  \\
       2 & 2 & 0 &  $G_{200}(e)$  \\
       2 & 2 & 1 &  $G_{20-1}(e)$ \\
       2 & 2 & 2 &  $G_{20-2}(e)$ \\
       2 & 2 & 3 &  $G_{20-3}(e)$ \\
       2 & 2 & 4 &  $G_{20-4}(e)$ \\
       2 & 2 & 5 &  $G_{20-5}(e)$ \\
       2 & 2 & 6 &  $G_{20-6}(e)$ \\
       2 & 2 & 7 &  $G_{20-7}(e)$ \\
       \hline  
         \end{tabular}
     \end{table}
       
    \end{appendix}

\end{document}